\documentclass{amsproc}
\usepackage{amsmath, amsthm, amssymb, slashed, framed}
\usepackage{ifpdf}
\ifpdf
  \usepackage[pdftex]{graphicx}
  \usepackage{epstopdf}
\else
  \usepackage[dvips]{graphicx}
\fi
\usepackage{url}
\newtheorem{theorem}{Theorem}[section]

\theoremstyle{definition}

\newtheorem{example}[theorem]{Example}

\newtheorem{conjecture}[theorem]{Conjecture}

\theoremstyle{remark}
\newtheorem{remark}[theorem]{Remark}

\newdimen\tableauside\tableauside=1.0ex
\newdimen\tableaurule\tableaurule=0.4pt
\newdimen\tableaustep
\def\phantomhrule#1{\hbox{\vbox to0pt{\hrule height\tableaurule width#1\vss}}}
\def\phantomvrule#1{\vbox{\hbox to0pt{\vrule width\tableaurule height#1\hss}}}
\def\sqr{\vbox{%
\phantomhrule\tableaustep
\hbox{\phantomvrule\tableaustep\kern\tableaustep\phantomvrule\tableaustep}%
\hbox{\vbox{\phantomhrule\tableauside}\kern-\tableaurule}}}
\def\squares#1{\hbox{\count0=#1\noindent\loop\sqr
\advance\count0 by-1 \ifnum\count0>0\repeat}}
\def\tableau#1{\vcenter{\offinterlineskip
\tableaustep=\tableauside\advance\tableaustep by-\tableaurule
\kern\normallineskip\hbox
    {\kern\normallineskip\vbox
      {\gettableau#1 0 }%
     \kern\normallineskip\kern\tableaurule}%
  \kern\normallineskip\kern\tableaurule}}
\def\gettableau#1 {\ifnum#1=0\let\next=\null\else
  \squares{#1}\let\next=\gettableau\fi\next}

\numberwithin{equation}{section}

\title{Homological algebra of knots and BPS states}

\author{Sergei Gukov}
\address{California Institute of Technology, Pasadena, CA 91125, USA\\
\newline
Max-Planck-Institut f\"ur Mathematik, Vivatsgasse 7, D-53111 Bonn, Germany.}
\email{gukov@theory.caltech.edu}
\author{Marko Sto$\check{\text{s}}$i$\acute{\text{c}}$}

\address{Instituto de Sistemas e Robotica and CAMGSD, Instituto Superior Tecnico, Torre
Norte, Piso 7, Av. Rovisco Pais, 1049-001 Lisbon, Portugal\\
\newline
Mathematical Institute SANU, Knez Mihailova 36, 11000 Beograd, Serbia.}
\email{mstosic@math.ist.utl.pt}

\thanks{%This is an extended write up of the Takagi Lectures, November 2010.Prepared for the Takagi Lectures 2010.
}

\font\teneurm=eurm10 \font\seveneurm=eurm7 \font\fiveeurm=eurm5
\newfam\eurmfam
\textfont\eurmfam=\teneurm \scriptfont\eurmfam=\seveneurm
\scriptscriptfont\eurmfam=\fiveeurm

 \font\teneusm=eusm10 \font\seveneusm=eusm7 \font\fiveeusm=eusm5
\newfam\eusmfam
\textfont\eusmfam=\teneusm \scriptfont\eusmfam=\seveneusm
\scriptscriptfont\eusmfam=\fiveeusm

\font\tencmmib=cmmib10 \skewchar\tencmmib='177
\font\sevencmmib=cmmib7 \skewchar\sevencmmib='177
\font\fivecmmib=cmmib5 \skewchar\fivecmmib='177
\newfam\cmmibfam
\textfont\cmmibfam=\tencmmib \scriptfont\cmmibfam=\sevencmmib
\scriptscriptfont\cmmibfam=\fivecmmib
\def\cmmib#1{{\fam\cmmibfam\relax#1}}

\def\example#1{\bgroup\narrower%\footnotefont
\baselineskip\footskip\bigbreak
\hrule\medskip\nobreak\noindent {\bf Example}. {\it #1\/}\par\nobreak}
\def\endexample{\medskip\nobreak\hrule\bigbreak\egroup}

\newcommand{\be}{\begin{equation}}
\newcommand{\ee}{\end{equation}}
\newcommand{\bea}{\begin{eqnarray}}
\newcommand{\eea}{\end{eqnarray}}

\newcommand{\C}{\mathbb{C}}
\newcommand{\Z}{\mathbb{Z}}
\newcommand{\R}{\mathbb{R}}
\newcommand{\Q}{\mathbb{Q}}

\newcommand{\g}{{\mathfrak g}}

\def\frak{\mathfrak}

\def\tilde{\widetilde}
\def\hat{\widehat}
\def\bar{\overline}

\def\CB{{\mathcal B}}
\def\CC{{\mathcal C}}

\def\CH{{\mathcal H}}

\def\CJ{{\mathcal J}}

\def\CO{{\mathcal O}}
\def\CP{{\mathcal P}}

\def\CW{{\mathcal W}}

\def\CZ{{\mathcal Z}}

\def\a{\cmmib a}

\def\unknot{{\,\raisebox{-.08cm}{\includegraphics[width=.4cm]{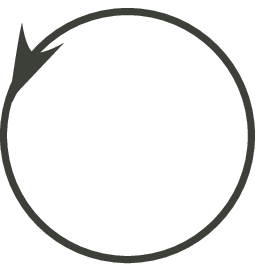}}\,}}

\def\cp{{\mathbb{C}}{\mathbf{P}}}

\def\leadsto{\rightsquigarrow}

\DeclareMathOperator{\rank}{rank}

\begin{document}

%    General info
\date{November, 2011}

\begin{abstract}
It is known that knot homologies admit a physical description as spaces of open BPS states.
We study operators and algebras acting on these spaces.
This leads to a very rich story, which involves wall crossing phenomena,
algebras of closed BPS states acting on spaces of open BPS states,
and deformations of Landau-Ginzburg models.

One important application to knot homologies is the existence of ``colored differentials''
that relate homological invariants of knots colored by different representations.
Based on this structure, we formulate a list of properties of the colored HOMFLY homology
that categorifies the colored HOMFLY polynomial.
By calculating the colored HOMFLY homology for symmetric and anti-symmetric representations,
we find a remarkable ``mirror symmetry'' between these triply-graded theories.
\end{abstract}

\maketitle

\tableofcontents
%\newpage
%\vspace*{20mm}

%%%%%%%%%%%%%%%%%%%%%%%%%%%%%%%%%%%%%%%%%%%%%%%%%%%%%%%%%%%%%%%%%%%%%%%%%%%%%%%%%%%%

\section{Setting the stage}
\label{sec:intro}

Quantum knot invariants were introduced in 1980's \cite{RT,Witt}:
for every representation $R$ of a Lie algebra $\g$, one can define a polynomial invariant $\bar{P}^{\g,R}(K)$ of a knot $K$.
Its {\it reduced} version is
\be
{P}^{\g,R}(K) \; = \; \frac{\bar{P}^{\g,R}(K)}{\bar{P}^{\g,R}({\,\raisebox{-.08cm}{\includegraphics[width=.4cm]{unknot}}\,})} \,,
\ee
where ${\,\raisebox{-.08cm}{\includegraphics[width=.4cm]{unknot}}\,}$ denotes the unknot.

A categorification of the polynomial ${P}^{\g,R}(K)$ (or its {\it unreduced} version $\bar{P}^{\g,R}(K)$)
is a doubly-graded homology theory\footnote{All homologies in this paper are defined over $\Q$.}
${\CH}^{\g,R}(K)$ whose graded Euler characteristic is equal to ${P}^{\g,R}(K)$.
In other words, if ${\CP}^{\g,R}(K)(q,t)$ denotes the Poincar\'e polynomial of ${\CH}^{\g,R}(K)$, then we have
\[
P^{\g,R}(K)(q) \; = \; \CP^{\g,R}(K)(q,t=-1) \,.
\]
Unlike ${P}^{\g,R}(K)$, the explicit combinatorial definition of ${\CH}^{\g,R}(K)$ exists for very few choices of $\g$ and $R$.
However, physics insights based on BPS state counting and Landau-Ginzburg theories predict
various properties and a very rigid structure of these homology theories.

One of the first results was obtained in \cite{DGR} for $\g=sl(N)$ and its fundamental representation $R\!=\!\tableau1$.
This work builds on a physical realization of knot homologies as spaces of BPS states \cite{GSV,GukovRTN}:
\be
\CH_{\text{knot}} \; = \; \CH_{\text{BPS}} \,.
\label{HHBPS}
\ee
%\cite{GSV}, where a relation between homological invariants of knots and the so-called {\it refined open BPS invariants} was proposed.
Among other things, this relation predicts the existence of a polynomial knot invariant $\CP^{\tableau1}{(K)}(\a,q,t)$,
sometimes called the {\it{superpolynomial}}, such that for all sufficiently large $N$ one has
\be
{\CP}^{sl(N),\tableau1}(K)(q,t) \; = \; \CP^{\tableau1}(K)(\a=q^N,q,t) \,.
\label{maja2}
\ee
Moreover, the polynomial $\CP^{\tableau1}(K)(\a,q,t)$
has nonnegative coefficients and is equal to the Poincar\'e polynomial of a triply graded homology theory ${\CH}^{\tableau1}(K)$
that categorifies the reduced two-variable HOMFLY polynomial $P^{\tableau1}(K)(\a,q)$, and similarly for the unreduced invariants.
This triply graded theory comes equipped with a collection of differentials $\{d_N\}$,
such that the homology of ${\CH}^{\tableau1}(K)$ with respect to $d_N$ is isomorphic to ${\CH}^{sl(N),\tableau1}(K)$.

There are only two triply-graded knot homologies that have been studied in the literature up to now.
Besides the above-mentioned HOMFLY homology, the second triply-graded theory, proposed in \cite{GWalcher},
similarly unifies homological knot invariants for the $N$-dimensional vector representation $R=V$ of $\g = so(N)$ and $\g = sp(N)$.
This triply-graded theory $\CH^{\mathrm{Kauff}}(K)$ comes with a collection
of differentials $\{d_N\}$, such that the homology with respect to $d_N$ for $N>1$ is isomorphic to $\CH^{so(N),V}(K)$,
while the homology with respect to $d_N$ for even $N<0$ is isomorphic to $\CH^{sp(-N),V}(K)$.
Since the graded Euler characteristic of $\CH^{\mathrm{Kauff}}(K)$ is equal to the (reduced) Kauffman polynomial of $K$,
$\CH^{\mathrm{Kauff}}(K)$ is called the Kauffman homology of a knot $K$.

One way to discover differentials acting on all of these knot homology theories is via studying deformations of the potentials and matrix factorizations in the corresponding Landau-Ginzburg theories (see section \ref{sec:matrix} for details).
In particular, in the case of the Kauffman homology one finds a peculiar deformation that leads to a ``universal" differential $d_{\to}$
and its conjugate $d_{\leftarrow}$, such that the homology with respect to these differentials is, in both cases,
isomorphic to the triply-graded HOMFLY homology $\CH^{\tableau1}(K)$.\\

A careful reader may notice that most of the existent results reviewed here deal with the fundamental
or vector representations of classical Lie algebras (of Cartan type $A$, $B$, $C$, or $D$).
In this paper, we do roughly the opposite: we focus mainly on $\g = sl(N)$ but vary the representation $R$.
In particular, we propose infinitely many triply-graded homology theories associated with arbitrary symmetric ($S^r$)
and anti-symmetric ($\Lambda^r$) representation of $sl(N)$.
Moreover, these colored HOMFLY homology theories come equipped with differentials,
such that the homology, say, with respect to $d_N^{S^r}$ is isomorphic to $\CH^{sl(N),S^r}(K)$, and similarly for $R = \Lambda^r$.

Remarkably, in addition to the differentials labeled by $N$ (for a given $r$) we also find {\it{colored}} differentials
that allow to pass from one triply graded theory to another, thus relating homological knot invariants associated with different representations!

Specifically, for each pair of positive integers $(r,m)$ with $r>m$, we find a differential $d_{r\to m}$,
such that the homology of $\CH^{S^r}(K)$ with respect to $d_{r\to m}$ is isomorphic to $\CH^{S^m}(K)$.
Similarly, in the case of anti-symmetric representations,
we find an infinite sequence of triply-graded knot homologies $\CH^{\Lambda^r}(K)$, one for every positive integer $r$,
equipped with colored differentials that allow to pass between two triply-graded theories with different values of $r$.\\

The colored differentials are a part of a larger algebraic structure that becomes manifest
in a physical realization of knot homologies as spaces of BPS states.
As it often happens in physics, the same physical system may admit several mathematical descriptions;
a prominent example is the relation between Donaldson-Witten and Seiberg-Witten invariants of 4-manifolds
that follows from physics of supersymmetric gauge theories in four dimensions \cite{Wmonopoles}.
Similarly, the space of BPS states in \eqref{HHBPS} admits several (equivalent) descriptions
depending on how one looks at the system of five-branes in eleven-dimensional M-theory \cite{GSV} relevant to this problem.

Specifically, for knots in a 3-sphere ${\bf S}^3$
the relevant system is a certain configuration of five-branes in M-theory on $\R \times M_4 \times X$,
where $M_4 \cong \R^4$ is a 4-manifold with isometry group $U(1)_P \times U(1)_F$ and $X$ is a non-compact toric Calabi-Yau 3-fold
(both of which will be discussed below in more detail).
And, if one looks at this M-theory setup from the vantage point of the Calabi-Yau space $X$,
one finds a description of BPS states via enumerative geometry of $X$.
Furthermore, for simple knots and links that preserve toric symmetry of the Calabi-Yau 3-fold $X$
the study of enumerative invariants reduces to a combinatorial problem of counting certain 3d partitions
(= fixed points of the 3-torus action \cite{ORV}),
hence, providing a combinatorial formulation of knot homologies in terms of 3d partitions~\cite{GIKV,IK}.

On the other hand, if one looks at this M-theory setup from the vantage point of the 4-manifold $M_4$,
one can express the counting of BPS invariants in terms of equivariant instanton counting on $M_4$.
In this approach (see {\it e.g.} \cite{DGH}), the ``quantum'' $q$-grading and the homological $t$-grading
on the space \eqref{HHBPS} originate from the equivariant action of $U(1)_P \times U(1)_F$ on $M_4$.

A closely related viewpoint, that will be very useful to us in what follows,
is based on the five-brane world-volume theory \cite{fiveknots}.
Let us briefly review the basic ingredients of this approach
that will make the relation to the setup of \cite{GSV} more apparent.
In both cases, knot homology is realized as the space of BPS states and, as we shall see
momentarily, the physical realization of the triply-graded knot homology proposed in \cite{GSV}
is essentially the large-$N$ dual of the system realizing the doubly-graded knot homology in \cite{fiveknots}.
This is very typical for systems with $SU(N)$ gauge symmetry\footnote{The same is true for other classical groups.}
which often admit a dual ``holographic'' description
that comprises all $N$ in the same package and leads to useful computational techniques~\cite{holography}.

In the case of $sl(N)$ homological knot invariants,
the five-brane configuration described in \cite[sec. 6]{fiveknots} is the following:
\bea
\text{space-time} & : & \qquad \R \times T^* W \times M_4 \nonumber \\
N~\text{M5-branes} & : & \qquad \R \times W \times D \label{theoryA} \\
\text{M5-brane} & : &  \qquad \R \times L_K \times D \nonumber
\eea
Here, $W$ is a 3-manifold and $D \cong \R^2$ is the ``cigar'' in the Taub-NUT space $M_4 \cong \R^4$.
The Lagrangian submanifold $L_K \subset T^* W$ is the conormal bundle to the knot $K \subset W$; in particular,
\be
L_K \cap W = K \,.
\ee
In all our applications, we consider $W = {\bf S}^3$ (or, a closely related case of $W = \R^3$).
Similarly, the setup of \cite{GSV} can be summarized as
\bea
\text{space-time} & : & \qquad \R \times X \times M_4 \label{theoryB} \\
\text{M5-brane} & : &  \qquad \R \times L_K \times D \nonumber
\eea
where $X$ is the resolved conifold, {\it i.e.} the total space of the $\CO (-1) \oplus \CO (-1)$ bundle over $\cp^1$.
{}From the way we summarized \eqref{theoryA} and \eqref{theoryB}, it is clear that they
have many identical elements.
%In particular, in both cases, knot homology is realized as the space of BPS states \eqref{HHBPS}.
The only difference is that \eqref{theoryA} has extra M5-branes supported on $\R \times W \times D$,
whereas \eqref{theoryB} has a different space-time (with a 2-cycle in the Calabi-Yau 3-fold $X$),
which is exactly what one expects from a holographic duality or large-$N$ transition \cite{GopakumarV,OoguriV}.

Indeed, what is important for the purpose of studying the space of BPS states, $\CH_{\text{BPS}}$, is that
both \eqref{theoryA} and \eqref{theoryB} preserve the same amount of supersymmetry and have the same symmetries:

\begin{itemize}

\item {\bf time translations:} both systems have a translation symmetry along the time direction
(denoted by the $\R$ factor in \eqref{theoryA} and \eqref{theoryB}).
Therefore, in both cases, one can ask for a space of BPS-states --- on multiple M5-branes in \eqref{theoryA},
and on a single M5-brane in \eqref{theoryB} --- which is precisely what was proposed as a candidate
for the $sl(N)$ knot homology (resp. HOMFLY homology).

\item {\bf rotation symmetries:}
\be
U(1)_P \times U(1)_F
\label{uusyms}
\ee
Here, the two $U(1)$ factors correspond, respectively, to the rotation symmetry
of the tangent and normal bundle of $D \cong \R^2$ in a 4-manifold $M_4 \cong \R^4$.
In particular, in both frameworks \eqref{theoryA} and \eqref{theoryB},
the former is responsible for the $q$-grading of $\CH_{BPS}$,
which corresponds to the conserved angular momentum derived from the rotation symmetry of $D$.

\end{itemize}

\noindent
A well-known feature of the large-$N$ duality is that the rank of the gauge group
turns into a geometric parameter of the dual system ({\it cf.} \cite{holography} or \cite{GopakumarV}).
In the present case, it is the K\"ahler modulus of the Calabi-Yau 3-fold $X = \CO_{\cp^1} (-1) \oplus \CO_{\cp^1} (-1)$:
\be
N \; \sim \; \log (\a) \; = \; {\rm Vol} (\cp^1) \,.
\label{NvolCP1}
\ee
The reason we denote the K\"ahler parameter by $\log (\a)$ rather than $\a$
is that with this convention $\a = q^N$ is the standard variable of the HOMFLY polynomial / knot homology.

\begin{figure}[htb]
\centering
\includegraphics[width=3in]{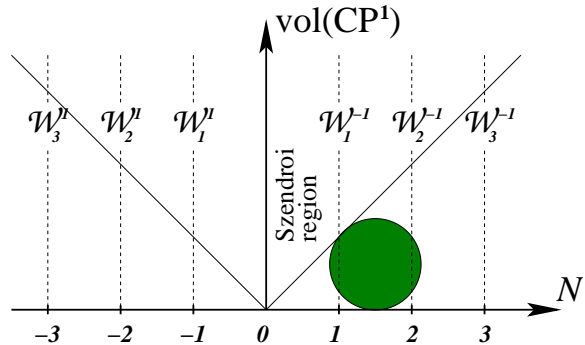}
\caption{For the conifold $X$, the space of stability conditions is one-dimensional.
It is parametrized by the K\"ahler parameter $N \sim \log (a) = {\rm Vol} (\cp^1)$.
This space is divided by walls of marginal stability into a set of chambers,
which can be identified with the set of integers $\Z$.}
\label{fig:walls}
\end{figure}

Another feature familiar to the practitioners of the refined / motivic Donaldson-Thomas theory
is that $\CH_{\text{BPS}}$ can jump as one varies stability conditions \cite{DenefM,KS1,Refined,BBS,CecottiV,DGS}.
Thus, in a closely related type IIA superstring compactification on a Calabi-Yau 3-fold~$X$,
the stability parameters are the K\"ahler moduli of $X$,
and in the present case there is only one K\"ahler modulus \eqref{NvolCP1} given by the volume of the $\cp^1$ cycle in $X$.
Therefore, we conclude that the space \eqref{HHBPS} can jump as one changes the stability parameter $N \sim {\rm Vol} (\cp^1)$.
%Moreover, the fact that only the refined BPS invariants jump implies that BPS states can appear and disappear only in pairs
%of states, related by the action of a supercharge $Q_{BRST}$.

Luckily, in the case where $X$ is the total space of the $\CO (-1) \oplus \CO (-1)$ bundle over $\cp^1$ relevant to
our applications, the wall-crossing behavior of the refined BPS invariants has been studied in the literature \cite{NagaoN,JafferisM,Refined}.
The one-dimensional space of stability conditions is divided into a set of chambers illustrated in Figure~\ref{fig:walls}.
In each chamber, $\CH_{\text{BPS}}$ is constant and the jumps of closed BPS states
occur at the walls $\CW_{n}^{\pm 1}$ characterized by different types of ``fragments'':
\be
\begin{array}{ccl}
\CW_n^1 & : & \quad D2/D0 \text{ fragments} \;  \\[.1cm]
\CW_n^{-1} & : & \quad \overline{D2}/D0 \text{ fragments} \;  \\[.1cm]
\CW_n^0 & : & \quad D0 \text{ fragments}
\end{array}\label{walltypes}
\ee
Notice, the set of chambers in this model can be identified with $\Z$, the set of integer numbers.
As we explain in the next section, this is not a coincidence. Namely, as we shall see,
every fragment corresponds to a differential acting on the space \eqref{HHBPS},
%so that in the present example one finds a set of differentials, $d_N$, labeled by~$N \in \Z$.
so that in the present example one finds a set of differentials $\{d_N\}$ labeled by~$N \in \Z$.

%%%%%%%%%%%%%%%%%%%%%%%%%%%%%%%%%%%%%%%%%%%%%%%%%%%%%%%%%%%%%%%%%%%%%%%%%%%%%%%%%%%%

The differentials $\{d_N\}$ are part of the homological algebra of knots / BPS states,
depending on whether one prefers to focus on the left or right side of the relation \eqref{HHBPS}.
For larger representations, in addition to the differentials $\{d_N\}$ one finds colored
differentials that allow to pass between homology theories associated with different $R$.
Even though a combinatorial definition of the majority of such theories, with all the differentials, is still missing,
their structure (deduced from physics) is so rigid that enables computation of
the homology groups for many knots and passes a large number of consistency checks.\\

In particular, by computing the triply-graded homologies $\CH^{S^r}(K)$ and $\CH^{\Lambda^r}(K)$ for various knots,
we find the following surprising symmetry between the two theories:
\be\label{mirsym}
\CH^{\Lambda^r}_{i,j,*}(K) \; \cong \; \CH^{S^r}_{i,-j,*}(K) \,.
\ee
One of the implication is that $\CH^{S^r}(K)$ and $\CH^{\Lambda^r}(K)$ can be combined into a single homology theory!

\begin{conjecture}\label{con0}
For every positive integer $r$, there exists a triply-graded theory $\CH^r(K)$ together with a collection of differentials $\{d^r_N\}$,
with $N\in \Z$, such that the homology of $\CH^r(K)$ with respect to $d^r_N$, for $N>0$, is isomorphic to $\CH^{sl(N),S^r}(K)$,
while the homology of $\CH^r(K)$ with respect to $d^r_N$, for $N<0$, is isomorphic (up to a simple regrading) to $\CH^{sl(-N),\Lambda^r}(K)$.
\end{conjecture}

Moreover, it is tempting to speculate that the symmetry \eqref{mirsym} extends to all representations:
\be\label{mirwild}
\text{``mirror symmetry''} ~:~ \quad
\CH^{\lambda} (K) \; \cong \; \CH^{\lambda^t} (K) \,,
\ee
where $\lambda$ and $\lambda^t$ are a pair of Young tableaux related by transposition (mirror reflection across the diagonal),
{\it e.g.}
$$
\lambda \; = \; \tableau{6 3 1} \qquad \longleftrightarrow \qquad \lambda^t \; = \; \tableau{3 2 2 1 1 1}
$$
The symmetry \eqref{mirwild} has not been discussed in physical or mathematical literature before.

While we offer its interpretation in section \ref{sec:physmir},
we believe the mirror symmetry for colored knot homology \eqref{mirwild}
deserves a more careful study, both in physics as well as in mathematics.
In particular, its deeper understanding should lead
to the ``categorification of level-rank duality'' in Chern-Simons theory,
which is the origin of the simpler, decategorified version of \eqref{mirwild}:
\be\label{mirHOMFLY}
P^{\lambda}(K)(\a,q) \; = \; P^{\lambda^t}(K)(\a,q^{-1})
\ee
for colored HOMFLY polynomials \cite{NaculichSchnitzer,Naculich,LMV,LiuPeng},
and extends the familiar symmetry $q \leftrightarrow q^{-1}$ of the ordinary HOMFLY polynomial.
We plan to pursue the categorification of level-rank duality
and to study the new, homological symmetry \eqref{mirwild} in the future work.

\subsection*{Organization of the paper}
We start by explaining in section \ref{sec:algebra} that, in general, the space of open BPS states forms a representation
of the algebra of closed BPS states. Then, in section \ref{sec:matrix} we review elements of the connection between
string realizations \eqref{theoryA}--\eqref{theoryB} of knot homologies and Landau-Ginzburg models that play an important role
in mathematical formulations of certain knot homologies based on Lie algebra $\g$ and its representation $R$.
In particular, we illustrate in simple examples how the corresponding potentials $W_{\g,R}$
can be derived from the physical setup \eqref{theoryA}--\eqref{theoryB} and how deformations of these potentials
lead to various differentials acting on ${\CH}^{\g,R}(K)$. This gives another way to look at the algebra acting on \eqref{HHBPS}.
Based on these predictions, in section \ref{sec:reduced} we summarize the mathematical structure
of the triply-graded homology $\CH^{S^r}(K)$, together with its computation for small knots.
Section \ref{sec:mirror} lists the analogous properties of the homology associated with anti-symmetric representations,
and explains the explicit form of the ``mirror symmetry" \eqref{mirsym} between symmetric and anti-symmetric triply-graded theories.
Unreduced triply-graded theory for symmetric and anti-symmetric representations is briefly discussed in section \ref{sec:unreduced}.
In appendix \ref{sec:notation} we collect the list of our notations,
whereas in appendix \ref{sec:942} we present the computations of the $S^2$, $\Lambda^2$ and Kauffman triply-graded homology
for knots $8_{19}$ and $9_{42}$.
These particular examples of ``thick'' knots provide highly non-trivial tests of all the properties of the homologies presented in the paper. Appendix \ref{apb41} contains the computation of the $S^3$ and $\Lambda^3$ homology of the figure-eight knot $4_1$.
Finally, appendix \ref{compun} collects some notations and calculations relevant to the unreduced
colored HOMFLY polynomial of the unknot discussed in section \ref{sec:unreduced}.

\section{Algebra of BPS states and its representations}
\label{sec:algebra}

Differentials in knot homology form a part of a larger
algebraic structure that has an elegant interpretation in the geometric / physical framework.
Because this algebraic structure has analogs in more general string / M-theory compactifications,
in this section we shall consider aspects of such structure for an arbitrary Calabi-Yau 3-fold $X$
with extra branes supported on a general Lagrangian submanifold $L \subset X$, {\it e.g.}
\bea
\text{space-time} & : & \qquad \R \times X \times M_4 \label{theoryC} \\
\text{M5-brane} & : &  \qquad \R \times L \times D \nonumber
\eea
For applications to knot homologies, one should take $X$ to be the total space of the $\CO (-1) \oplus \CO (-1)$ bundle over $\cp^1$
and $L_K$ to be the Lagrangian submanifold determined by a knot $K$ \cite{OoguriV,Taubes,Koshkin}.
Then, \eqref{theoryC} becomes precisely the setup \eqref{theoryB},
in which homological knot invariants are realized as spaces of refined BPS states, {\it cf.} \eqref{HHBPS}.

In fact, there are {\it two} spaces of BPS states relevant to this particular problem
and its variants based on a more general 3-fold $X$.
One is the space of refined {\it closed} BPS states, denoted as $\CH_{\text{BPS}}^{\text{closed}}$,
and the other is called the space of refined {\it open} BPS states, $\CH_{\text{BPS}}^{\text{open}}$.
The difference is that, while the latter contains BPS particles in the presence
of defects\footnote{M5-branes in the M-theory setup \eqref{theoryC}},
the former comprises only those BPS states which are still present in a theory when all defects are removed.
In the description~\cite{DGH} via equivariant instanton counting on a 4-manifold $M_4$,
the defect (M5-brane) corresponds to a particular ramification along the divisor $D \subset M_4$,
the so-called surface operator.

On the other hand, if one looks at the general setup \eqref{theoryC} from the vantage point
of the Calabi-Yau space $X$, then $\CH_{\text{BPS}}^{\text{closed}}$ and $\CH_{\text{BPS}}^{\text{open}}$
can be formulated in terms of enumerative invariants of $X$ and $(X,L)$ that ``count'', respectively,
closed holomorphic curves embedded in $X$
and bordered holomorphic Riemann surfaces $(\Sigma, \partial \Sigma) \hookrightarrow (X,L)$
with boundary on the Lagrangian submanifold $L$. As a way to remember this, it is convenient to keep in mind that

\medskip

\begin{itemize}

\item
$\CH_{\text{BPS}}^{\text{closed}}$ depends only on the Calabi-Yau space $X$

\bigskip

\item
$\CH_{\text{BPS}}^{\text{open}}$ depends on both the Calabi-Yau space $X$ and the Lagrangian submanifold $L \subset X$

\end{itemize}

\medskip
\noindent
In applications to knots, open (resp. closed) BPS states are represented by open (resp. closed)
membranes in the M-theory setup \eqref{theoryB} or by bound states of D0 and D2 branes in its reduction to type IIA string theory.
It is the space of open BPS states that depends on the choice of the knot $K$ and, therefore,
provides a candidate for homological knot invariant in \eqref{HHBPS}.

In general, the space of BPS states is $\Gamma \oplus \Z$-graded, where $\Gamma$ is the ``charge lattice''
and the extra $\Z$-grading comes from the (half-integer) spin of BPS states, such that $2j_3 \in \Z$.
For example, in the case of closed BPS states, the charge lattice
is usually just the cohomology lattice of the corresponding Calabi-Yau 3-fold $X$,
\be
\Gamma \; = \; H^{\text{even}} (X;\Z) \,.
\ee
In the case of open BPS states, $\Gamma$ also depends on the choice of the Lagrangian submanifold $L \subset X$.

When $X$ is the total space of the $\CO (-1) \oplus \CO (-1)$ bundle over $\cp^1$ and $L = L_K$,
as in application to knot homologies, the lattice $\Gamma$ is two-dimensional for both open and closed BPS states.
As a result, both $\CH_{\text{BPS}}^{\text{closed}}$ and $\CH_{\text{BPS}}^{\text{open}}$
are $\Z \oplus \Z \oplus \Z$-graded.
In particular, the space of open BPS states
is graded by spin $2j_3 \in \Z$ and by charge $\gamma = (n,\beta) \in \Gamma$, where the degree $\beta \in H_2 (X,L_K) \cong \Z$
is sometimes called the ``D2-brane charge'' and $n \in \Z$ is the ``D0-brane charge.''
In relation to knot homologies \eqref{HHBPS}, these become the three gradings of the theory categorifying the colored HOMFLY polynomial:
\be
\begin{array}{rcl}
``\a-\text{grading}" & = & \beta \in H_2 (X,L_K) \cong \Z \;  \\[.2cm]
``q-\text{grading}" & = & n \in \Z \;  \\[.2cm]
``t-\text{grading}" & = & 2 j_3 \in \Z
\end{array}
\label{aqtgradings}
\ee
\\

Now, let us discuss the algebraic structure that will help us understand the origin of differentials
acting on the triply-graded vector space $\CH_{\text{knot}} = \CH_{\text{BPS}}^{\text{open}}$.
The fact that $\CH_{\text{BPS}}^{\text{closed}}$ forms an algebra is well appreciated
in physics \cite{HarveyMoore} as well as in math literature \cite{KSCOHA}.
Less appreciated, however, is the fact that $\CH_{\text{BPS}}^{\text{open}}$
forms a representation of the algebra $\CH_{\text{BPS}}^{\text{closed}}$:
\be
\boxed{\phantom{\int}
\begin{array}{rcl}
\text{refined open BPS states}: &  & \CH_{\text{BPS}}^{\text{open}} \\
 & & \circlearrowleft \\
\text{refined closed BPS states}: &  & \CH_{\text{BPS}}^{\text{closed}}
\end{array}
\phantom{\int}}
\label{HHrep}
\ee
Indeed, two closed BPS states, $\CB_1$ and $\CB_2$, of charge $\gamma_1, \gamma_2 \in \Gamma$
can form a bound state, $\CB_{12}$ of charge $\gamma_1 + \gamma_2$, as a sort of ``extension'' of $\CB_1$ and $\CB_2$,
\be
0 \; \to \; \CB_2 \; \to \; \CB_{12} \; \to \; \CB_1 \; \to \; 0 \,,
\label{extBBB}
\ee
thereby defining a product on $\CH_{\text{BPS}}^{\text{closed}}$:
%the vector space $\CH_{\text{BPS}}^{\text{closed}}$ comes with a product:
%\be
%\CH_{\text{BPS}}^{\text{closed}} \; \otimes \; \CH_{\text{BPS}}^{\text{closed}} \; \longrightarrow \; \CH_{\text{BPS}}^{\text{closed}}
%\ee
\be
\begin{array}{ccl}
\CH_{\text{BPS}}^{\text{closed}} \; \otimes \; \CH_{\text{BPS}}^{\text{closed}} & \longrightarrow & \CH_{\text{BPS}}^{\text{closed}} \;  \\[.4cm]
(\; \CB_1 \;, \;\; \CB_2 \;) & \mapsto & \CB_{12}
\end{array}
\label{clcl}
\ee
$$
{\,\raisebox{-.5cm}{\includegraphics[width=5cm]{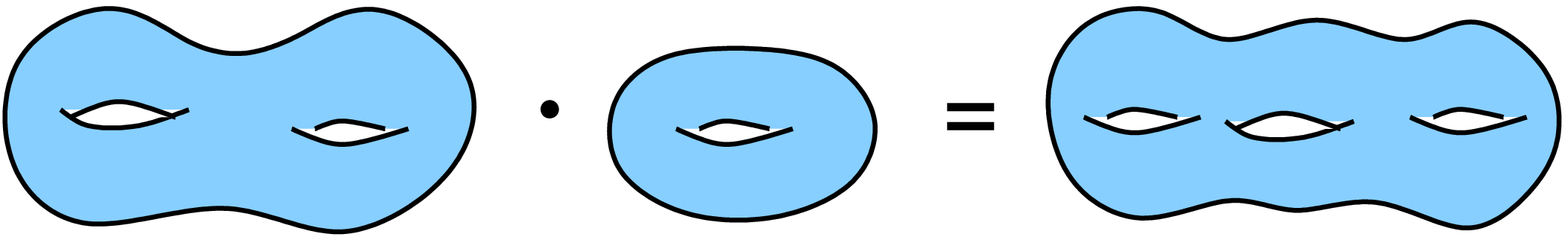}}\,}
$$
Similarly, a bound state of a closed BPS state $\CB_1^{\text{closed}} \in \CH_{\text{BPS}}^{\text{closed}}$
with an open BPS state $\CB_2^{\text{open}} \in \CH_{\text{BPS}}^{\text{open}}$
is another open BPS state $\CB_{12}^{\text{open}} \in \CH_{\text{BPS}}^{\text{open}}$:
\be
(\; \CB_1^{\text{closed}} \;, \;\; \CB_2^{\text{open}} \;) \quad \mapsto \quad \CB_{12}^{\text{open}}
\label{opclosed}
\ee
$$
{\,\raisebox{-.5cm}{\includegraphics[width=5cm]{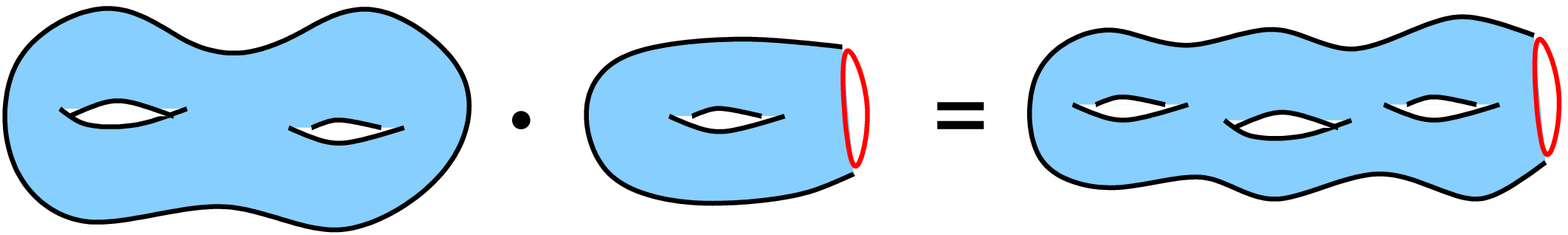}}\,}
$$
This defines an action of the algebra of closed BPS states on the space of open BPS states.

The process of formation or fragmentation of a bound state in \eqref{clcl} and \eqref{opclosed} takes place
when the binding energy vanishes.
Since the energy of a BPS state is given by the absolute value of the central charge\footnote{The central
charge function is a linear function in the sense that $\CZ (\gamma_1 + \gamma_2) = \CZ (\gamma_1) + \CZ (\gamma_2)$,
{\it i.e.} it defines a homomorphism $\CZ \in \text{Hom} (\Gamma, \C)$.}
function $\CZ : \Gamma \to \C$ this condition can be written as
\be
\big| \CZ (\gamma_1 + \gamma_2) \big|
- \big| \CZ (\gamma_1) \big| - \big| \CZ (\gamma_2) \big| \; = \; 0
\label{binding}
\ee
for a process that involves either $\CB_{12} \to \CB_1 + \CB_2$ or its inverse $\CB_1 + \CB_2 \to \CB_{12}$.
A particular instance of the relation \eqref{binding} is when the central charge of the fragment vanishes:
\be
\CZ (\gamma_{\text{fragment}}) \; = \; 0
\ee
Then, a fragment becomes massless and potentially can bind to any other BPS state of charge $\gamma$.
When combined with \eqref{HHrep},
it implies that closed BPS states of zero mass correspond to operators
acting on the space of open BPS states $\CH_{\text{BPS}}^{\text{open}}$.
The degree of the operator is determined by the spin and charge of the corresponding BPS state, as in \eqref{aqtgradings}.

For example, when $X$ is the total space of the $\CO (-1) \oplus \CO (-1)$ bundle over $\cp^1$,
as in application to knot homologies, we have
\be
\exp (\CZ) \; = \; \a^{\beta} q^n \,,
\ee
where we used the relation \eqref{NvolCP1} between $\a$ and ${\rm Vol} (\cp^1)$.
Therefore, for special values of $\a$ and $q$ we have the following massless fragments, {\it cf.} \eqref{walltypes}:
\be
\begin{array}{lll}
\a=q^{-N} & : & \quad D2/D0 \text{ fragments} \;  \\[.1cm]
\a=q^N & : & \quad \overline{D2}/D0 \text{ fragments} \;  \\[.1cm]
q=1 & : & \quad D0 \text{ fragments}
\end{array}\label{frogtypes}
\ee
Moreover, the D2/D0 fragments obey the Fermi-Dirac statistics (see {\it e.g.} \cite{Refined,JafferisM})
and, therefore, lead to anti-commuting operators ({\it i.e.} differentials) on $\CH_{\text{BPS}}^{\text{open}}$.

To summarize, we conclude that various specializations of the parameters (stability conditions)
are accompanied by the action of commuting and anti-commuting operators on $\CH_{\text{BPS}}^{\text{open}}$.
The algebra of these operators is precisely the algebra of closed BPS states $\CH_{\text{BPS}}^{\text{closed}}$.
Mathematical candidates for the algebra of closed BPS states include variants of the Hall algebra \cite{Schiffmann},
which by definition encodes the structure of the space of extensions \eqref{extBBB}:
\be
[\CB_1] \cdot [\CB_2] \; = \; \sum_{\CB_{12}} |0 \to \CB_2 \to \CB_{12} \to \CB_1 \to 0| \; [\CB_{12}]
\ee
In the present case, the relevant algebras include the motivic Hall algebra \cite{KS1},
the cohomological Hall algebra \cite{KSCOHA},
and its various ramifications, {\it e.g.} cluster algebras.
Therefore, the problem can be approached by studying representations of these algebras, as will be described elsewhere.

%%%%%%%%%%%%%%%%%%%%%%%%%%%%%%%%%%%%%%%%%%%%%%%%%%%%%%%%%%%%%%%%%%%%%

\section{B-model and matrix factorizations}
\label{sec:matrix}

Let us denote by $\CH^{\frak{g}, R}$ a homology theory of knots and links colored by a representation $R$ of the Lie algebra $\frak{g}$.
Many such homology theories can be constructed using categories of matrix factorizations \cite{KRa,KRfoam,KRb,Khov,KRso,MSV,Yonezawa,Wu}.
In this approach, one of the main ingredients is a polynomial function $W_{\frak{g},R}$ called the {\it potential},
associated to every segment of a link (or, more generally, of a tangle) away from crossings.
For example, for the fundamental representation of $\frak{g} = sl(N)$ the potential
is a function of a single variable,
\be
W_{sl(N), \tableau{1}} (x) \; = \; x^{N+1} \,.
\label{WslN}
\ee

In physics, matrix factorizations are known \cite{Kontsevich,Kli,Orlov,BHLS,HWalcher,BrunnerR}
to describe D-branes and topological defects
in Landau-Ginzburg models which, in the present context, are realized on the two-dimensional
part of the five-brane world-volume in \eqref{theoryA} or \eqref{theoryB}.
More precisely, it was advocated in \cite{GWalcher} that reduction of the M-theory configuration \eqref{theoryA}
on one of the directions in $D$ and a T-duality along the time direction
gives a configuration of intersecting D3-branes in type IIB string theory,
such that the effective two-dimensional theory on their common world-volume provides a physical
realization of the Landau-Ginzburg model that appears in the mathematical constructions.

In particular, this interpretation was used to deduce potentials $W_{\frak{g}, R}$ associated to
many Lie algebras and representations. Indeed, since away from crossings every segment of the knot $K$
is supposed to be described by a Landau-Ginzburg theory with potential $W_{\frak{g}, R}$,
we can approximate this local problem by taking $W = \R^3$ and $K = \R$ in \eqref{theoryA}.
Then, we also have $L_K = \R^3$ and the reduction (plus T-duality) of \eqref{theoryA} gives type IIB theory
in flat space-time with two sets of D3-branes supported on 4-dimensional hyperplanes in $\R^{10}$:
one set supported on $\R \times W$, and another supported on $\R \times L_K$.
The space of open strings between these two groups of D3-branes contains information about the potential $W_{\frak{g}, R}$.

\begin{figure}[htb]
\centering
\includegraphics[width=2in]{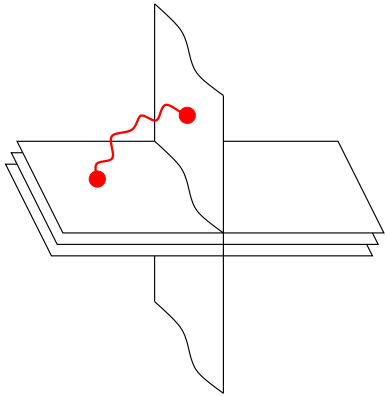}
\caption{The physics of open strings between two stacks of Lagrangian branes
is described by the Landau-Ginzburg model with potential~$W_{\frak{g}, R}$.}
\label{fig:branes}
\end{figure}

For example, in the case of the fundamental representation of $sl(N)$, the first stack consists of $N$ D3-branes
and the second only contains a single D3-brane. The open strings between these two stacks of D3-branes
transform in the bifundamental representation $({\bf N}, +1)$ under the gauge symmetry $U(N) \times U(1)$ on the D3-branes.
%(Here, the first factor describes gauge symmetry of $N$ D3-branes on $\R \times W$ and the second factor
%describes gauge symmetry of a single D3-brane supported on $\R \times L_K$.)
The Higgs branch of this two-dimensional theory
is the K\"ahler quotient of the vector space $\C^N$ parametrized by the bifundamental chiral multiplets,
modulo $U(1)$ gauge symmetry of a single D3-brane supported on $\R \times L_K$:
\be
\C^N / \! / U(1) \; \cong \; \cp^{N-1} \,.
\ee
The chiral ring of this theory on the intersection of D3-branes
is precisely the Jacobi ring of the potential \eqref{WslN}.

Following similar arguments one can find potentials associated to many other Lie algebras
and representations \cite{GWalcher}, such that
\be
\CH^{\frak{g}, R} (\unknot) \; = \; \CJ (W_{\frak{g}, R}) \,.
\ee
For example, the arguments that lead to \eqref{WslN} can be easily generalized to $R=\Lambda^r$,
the $r$-th anti-symmetric representation of $sl(N)$.
The only difference is that, in this case, the corresponding brane systems \eqref{theoryA} and \eqref{theoryB}
contain $r$ coincident M5-branes supported on $\R \times L_K \times D$.
Following the same arguments as in the case of the fundamental representation ($r=1$)
and zooming in closely on the local geometry of the brane intersection,
after all the dualities we end up with a system of intersecting D3-branes in flat ten-dimensional space-time,
\bea
N~\text{D3-branes} & : & \qquad \R \times W \label{D3branes} \\
r~\text{D3}'\text{-branes} & : &  \qquad \R \times L_K \nonumber
\eea
where, as in the previous discussion, for the purpose of deriving $W_{\frak{g},R}$
we can approximate $W \simeq \R^3$ and $L_K \simeq \R^3$, so that $W \cap L_K = \R$.
Now, the open strings between two sets of D3-branes in \eqref{D3branes}
transform in the bifundamental representation $({\bf N}, {\bf r})$ under the gauge symmetry $U(N) \times U(r)$ on the D3-branes.
Here, if we want to ``integrate out'' open strings ending on the D3$'$-branes,
only the second gauge factor should be considered dynamical,
while $U(N)$ should be treated as a global symmetry of the two-dimensional $U(r)$ gauge theory on the brane intersection.
In the infrared this theory flows to a sigma-model based on the Grassmannian manifold:
\be
{\rm Gr} (r,N) \; = \; \frac{U(N)}{U(r) \times U(N-r)} \,.
\ee
The potential of the corresponding Landau-Ginzburg model \cite{Wgrassmannian} is a homogeneous polynomial of degree $N+1$,
\be
W_{sl(N),\Lambda^r} (z_1,\ldots,z_r) \; = \; x^{N+1}_1 + \ldots + x_r^{N+1} \,,
\label{Wanti}
\ee
where the right-hand side should be viewed as a function of the variables $z_i$ of degree $\deg (z_i) = i$, $i=1, \ldots, r$,
which are the elementary symmetric polynomials in the $x_j$,
$$
z_i \; = \; \sum_{j_1<j_2<\cdots<j_i} x_{j_1}x_{j_2} \cdots x_{j_i} \,.
$$
We shall return to the discussion of the potential $W_{sl(N),\Lambda^r}$ later in this section.
In the case of more general representations,
one needs to consider various sectors of the $U(r)$ gauge theory on $\R \times L_K$
labeled by non-trivial flat connections (Wilson lines) around the codimension-2 locus
where D3$'$-branes meet D3-branes, {\it cf.}~\cite{OoguriV,Rigid}.

In this paper we are mostly interested in knots colored by symmetric and anti-symmetric representations of $\frak{g} = sl(N)$,
even though much of the present discussion can be easily generalized to other Lie algebras and representations.
Thus, for a symmetric representation $R=S^r$ of $\frak{g} = sl(N)$ one finds that
the corresponding potential $W_{sl(N),S^r} (z_1, \ldots, z_r)$ is a homogeneous polynomial of degree $N+r$
in variables $z_i$ of degree $i=1,\ldots,r$, much like \eqref{Wanti}.
Moreover, the explicit form of such potentials can be conveniently expressed through a generating function \cite{GWalcher}:
\be
\sum_N (-1)^{N} t^{N+r} W_{sl(N),S^r} (z_1, \ldots, z_r) \; = \; (1+ \sum_{i=1}^r t^i z_i) \log ( 1 + \sum_{i=1}^r t^i z_i) \,,
\label{Wsymgenfncn}
\ee
which in the basic case $N=r=2$ gives
\be
W_{sl(2),\tableau{2}} \; = \; z_1^4 - 6 z_1^2 z_2 + 6 z_2^2 \,.
\label{W22}
\ee
Instead of going through the derivation of this formula we can use a simple trick
based on the well known isomorphism $sl(2) \cong so(3)$
under which a vector representation of $so(3)$ is identified with the adjoint representation of $sl(2)$.
Indeed, it implies that \eqref{W22} should be identical to the well known potential
\be
W_{so(3), V} \; = \; x^2 + xy^2
\label{Wso3}
\ee
in the $so(3)$ homology theory, {\it cf.} \cite{GWalcher,KRso}.
It is easy to verify that the potentials \eqref{W22} and \eqref{Wso3} are indeed related
by a simple change of variables.\footnote{Explicitly, the change of variables that relates
\eqref{W22} and \eqref{Wso3} is given by:
\begin{eqnarray}
x & = & - \frac{1}{2} (\sqrt{2} + \sqrt{6}) \; z_1^2 + \sqrt{6} \; z_2 \nonumber \\
y & = & 2^{1/4} \; z_1 \nonumber
\end{eqnarray}
}

Moreover, the fact that the adjoint representation of $sl(2)$ is identical to the vector representation of $so(3)$
implies that
\be
\boxed{\phantom{\int}
\CH^{sl(2), \tableau{2}} (K) \; \cong \; \CH^{so(3),V} (K) \phantom{\int}}
\label{sl2so3}
\ee
should hold for every knot $K$.
In particular, it should hold for the unknot.
And, since $\CH^{so(3),V} (\unknot)$ is 3-dimensional,
it follows
\be
\dim \CH^{sl(2), \tableau{2}} (\unknot) \; = \; \dim \CH^{so(3),V} (\unknot) \; = \; 3 \,.
\label{HHunknot3dim}
\ee
This is indeed what one finds in physical realizations of knot homologies reviewed in section \ref{sec:intro}.
In the framework of \cite{GSV} the colored homology $\CH^{sl(2), \tableau{2}}$ of the unknot
was computed in \cite{GIKV} using localization with respect to the toric symmetry of the Calabi-Yau space $X$.
Similarly, in the gauge theory framework \cite{fiveknots} the moduli space of solutions on $\R^2$
with a single defect operator in the adjoint representation of the gauge group $G=SU(2)$
is the weighted projective space $\mathbb{W} \cp^2_{(1,1,2)}$ (= the space of Hecke modifications \cite{KW}, see also \cite{GWopers}).
In this approach, the colored homology $\CH^{sl(2), \tableau{2}} (\unknot)$
is given by the $L^2$ cohomology of the moduli space $\mathbb{W} \cp^2_{(1,1,2)}$
which is 3-dimensional, in agreement with \eqref{HHunknot3dim}.

\subsection{Colored differentials}

One of the reasons why we carefully reviewed the properties of the potentials $W_{\frak{g}, R}$
is that they hold a key to understanding the colored differentials.
Namely, in doubly-graded knot homologies constructed from matrix factorizations differentials
that relate different theories are in one-to-one correspondence with deformations
of the potentials \cite{Lee,BarNatan,Turner,KhovanovFr,Gornik,GWalcher}:
\be
\boxed{\phantom{\int} \text{differentials on} ~ \CH^{\frak{g}, R} \quad \Longleftrightarrow \quad \text{deformations of} ~W_{\frak{g}, R} \phantom{\int}}
\label{defdiff}
\ee
For example, deformations of the potential \eqref{WslN} of the form $\Delta W = \beta x^{M+1}$ with $M<N$,
correspond to differentials $d_M$ that relate $sl(N)$ and $sl(M)$ knot homologies (with $R = \tableau{1}$).

More generally, one can consider deformations $\Delta W$ of the potential $W_{\frak{g}, R}$
such that $\deg \Delta W < \deg W_{\frak{g}, R}$ and
\be
W_{\frak{g}, R} + \Delta W \; \simeq \; W_{\frak{g}', R'}
\label{Wdeformed}
\ee
for some Lie algebra $\frak{g}'$ and its representation $R'$.
Here, the symbol ``$\simeq$'' means that the critical point(s) of the deformed potential
is locally described by the new potential $W_{\frak{g}', R'}$.
A deformation of this form leads to a spectral sequence that relates knot
homologies $\CH^{\frak{g}, R}$ and $\CH^{\frak{g}', R'}$.
With the additional assumption that the spectral sequence converges after the first page one arrives at \eqref{defdiff}.
Moreover, the difference
\be
\deg W_{\frak{g}, R}-\deg \Delta W
\label{WWqgrading}
\ee
gives the $q$-grading of the corresponding differential.
Notice, the condition $\deg \Delta W < \deg W_{\frak{g}, R}$
implies that this $q$-grading is positive.

For example, among deformations of the degree-4 potential \eqref{W22} one finds
$\Delta W = z_1^3$, which leads to a differential of $q$-degree 1
that relates $\CH^{sl(2),\tableau{2}}$ and $\CH^{sl(2),\tableau{1}}$.
This deformation has an obvious analog for higher rank $S^2$-colored homology;
it deforms the homogeneous polynomial $W_{sl(N),\tableau{2}} (z_1,z_2)$ of degree $N+2$
in such a way that the deformed potential has a critical point described by the potential
$W_{\frak{g}', R'} = W_{sl(N),\tableau{1}} (z_1)$ of degree $N+1$.
Therefore, it leads to a colored differential of $q$-degree 1, such that
\be
\big( \CH^{sl(N),\tableau{2}} , \; d_{\text{colored}} \big) \; \cong \; \CH^{sl(N),\tableau{1}} \,.
\label{coloredHdH}
\ee
In section \ref{sec:reduced} we present further evidence for the existence of a differential
with such properties not only in the doubly-graded $sl(N)$ theory but also in the triply-graded
knot homology that categorifies the colored HOMFLY polynomial.\\

Similar colored differentials exist in other knot homologies associated with more general Lie algebras and representations.
Basically, a knot homology associated to a representation $R$ of the Lie algebra $\frak{g}$
comes equipped with a set of colored differentials that, when acting on $\CH^{\frak{g}, R}$,
lead to homological invariants associated with smaller representations (and, possibly, Lie algebras),
\be
\dim R' \; < \; \dim R \,.
\label{coloredRR}
\ee
While it would be interesting to perform a systematic classification of such colored differentials
using the general principle \eqref{defdiff},
in this paper we limit ourselves only to symmetric and anti-symmetric representations of $\frak{g} = sl(N)$.

As we already discussed earlier,
when $R=\Lambda^r$ is the $r$-th anti-symmetric representation of $sl(N)$
the corresponding Landau-Ginzburg potential \eqref{Wanti} is a homogeneous polynomial of degree $N+1$.
Equivalently, the potentials with a fixed value of $r$ can be organized into a generating function,
analogous to \eqref{Wsymgenfncn}:
\be
\sum_N (-1)^{N} t^{N+1} W_{sl(N),\Lambda^r} (z_1, \ldots, z_r) \; = \; \log ( 1 + \sum_{i=1}^r t^i z_i) \,.
\label{WWanti}
\ee
For example, in the first non-trivial case of $r=2$ there are only two variables,
$z_1 = x_1 + x_2$ and $z_2 = x_1 x_2$.
For $N=2$ one finds a ``trivial'' potential $W_{sl(2),\tableau{1 1}}$ of degree 3,
which corresponds to the fact that the anti-symmetric representation $R = \Lambda^2$ (also denoted $R = \tableau{1 1}$) is trivial in $sl(2)$.
For $N=3$, the existence of the anti-symmetric tensor $\epsilon^{ijk}$ identifies the second anti-symmetric representation $R = \tableau{1 1}$
with the fundamental representation of $sl(3)$.
The next case in this sequence, $N=4$, is the first example where the second anti-symmetric representation is not related to
any other representation of $sl(4)$.
According to \eqref{Wanti} and \eqref{WWanti}, the corresponding potential is a homogeneous polynomial of degree 5,
\be
W_{sl(4), \tableau{1 1}} \; = \; \frac{z_1^5}{5} - z_1^3 z_2 + z_1 z_2^2 \,.
\ee
Before studying deformations of this potential, we note that by a simple change of variables
it is related to the potential
\be
W_{so(6),V} \; = \; x^5 + x y^2
\ee
associated to a vector representation of $so(6)$. This is a manifestation of the well known isomorphism $sl(4) \cong so(6)$
under which the six-dimensional anti-symmetric representation $R = \tableau{1 1}$ of $sl(4)$ is identified with the vector representation of $so(6)$.
This isomorphism can help us understand deformations of the potential $W_{sl(4), \tableau{1 1}} = W_{so(6),V}$.
Indeed, the deformations of $W_{so(N),V}$ were already studied in \cite{GWalcher}; they include several deformations
which lead to {\it canceling} differentials and a deformation by $\Delta W = y^2$ that leads to a {\it universal} differential
$\CH^{so(N),V} \leadsto \CH^{sl(N-2),\tableau{1}}$.

In view of the relation $W_{sl(4), \tableau{1 1}} (z_1,z_2) = W_{so(6),V} (x,y)$, these deformations (and the corresponding differentials)
should be present in the $sl(4)$ theory as well. In particular, there are deformations of $W_{sl(4), \tableau{1 1}}$
that lead to canceling differentials and, more importantly, there is a deformation by $\Delta W = y^2$
that leads to the universal differential which relates $\CH^{so(6),V} \cong \CH^{sl(4),\tableau{1 1}}$ and $\CH^{sl(4),\tableau{1}}$.
Note, from the viewpoint of the $sl(4)$ knot homology, this is exactly the colored differential $d_{\text{colored}}$
that does not change the rank of the Lie algebra, but changes the representation.
Making use of \eqref{WWanti} it is easy to verify that, for all values of $N$,
the potential $W_{sl(N), \tableau{1 1}}$ admits a deformation by terms of degree $N$
that leads to $W_{sl(N), \tableau{1}}$ and, therefore, to the analog of \eqref{coloredHdH}:
\be
\big( \CH^{sl(N),\tableau{1 1}} , \; d_{\text{colored}} \big) \; \cong \; \CH^{sl(N),\tableau{1}} \,.
\label{coloredHdHanti}
\ee
Much like in the case of the symmetric representations, this colored differential as well as canceling
differentials come from the triply-graded theory that categorifies the $\tableau{1 1}$-colored HOMFLY polynomial
(see section \ref{sec:mirror} for details).

%%%%%%%%%%%%%%%%%%%%%%%%%%%%%%%%%%%%%%%%%%%%%%%%%%%%%%%%%%%%%%%%%%%%%

\section{Colored HOMFLY homology}
\label{sec:reduced}

In this section we propose structural properties of the triply-graded theory categorifying
the colored version of the reduced HOMFLY polynomial.
The central role in this intricate network of structural properties belongs to the colored differentials,
whose existence we already motivated in the previous sections.

\subsection{Structural properties}
Let $N$ and $r$  be positive integers, and
let $\CH^{sl(N),S^r}(K)$ denote a reduced doubly-graded homology theory categorifying $P^{S^r}_N(K)$,
the polynomial invariant of a knot $K$ labeled with the $r$-th symmetric representation of $sl(N)$.
$\CP^{S^r}_N$ denotes the Poincar\'e polynomial of $\CH^{sl(N),S^r}(K)$.
Motivated by physics, we expect that such theories with a given value of $r$ have a lot in common.

\begin{conjecture}\label{con1}
For a knot $K$ and a positive integer $r$, there exists a finite polynomial
$\CP^{S^r}(K)\in\Z_+[\a^{\pm 1},q^{\pm 1},t^{\pm 1}]$ such that
\begin{equation}\label{coneq1}
\CP^{S^r}_N(K)(q,t)=\CP^{S^r}(K)(\a=q^N,q,t),
\end{equation}
for all sufficiently large $N$.
\end{conjecture}

Since the left-hand side of (\ref{coneq1}) is a Poincar\'e polynomial of a homology theory,
all coefficients of $\CP^{S^r}(K)(\a,q,t)$ must be nonnegative.
This suggests that there exists a triply-graded homology theory whose Poincar\'e polynomial is equal to $\CP^{S^r}(K)(\a,q,t)$,
and whose Euler characteristic is equal to the normalized $S^r$-colored two-variable HOMFLY polynomial.

As in the case of ordinary HOMFLY homology \cite{DGR} (that, in fact, corresponds to $r=1$) and in the case of Kauffman homology \cite{GWalcher},
this triply-graded theory comes with the additional structure of differentials, that will imply Conjecture \ref{con1}.
In particular, for each positive integer $r$ we have a triply-graded homology theory of a knot $K$.
Moreover, these theories come with additional structure of differentials that, as in \eqref{coloredRR},
allow us to pass from the homology theory with $R = S^r$ to theories with $R' = S^m$ and $m<r$.\\

Thus, we arrive to our main conjecture that describes the structure of the triply-graded homology
categorifying the $S^r$-colored HOMFLY polynomials:

\begin{conjecture}\label{con2}
For every positive integer $r$ there exists a triply-graded homology theory $\CH^{S^r}_*=\CH^{S^r}_{i,j,k}(K)$
that categorifies the reduced two-variable $S^r$-colored HOMFLY polynomial of $K$.
It comes with a family of differentials $\{d^{S^r}_N\}$, with $N\in\Z$, and also with an additional collection of \textit{universal colored} differentials $d_{r\to m}$, for every $1\le m<r$, satisfying the following properties:\\

\noindent$\bullet$ \textbf{Categorification: } $\CH^{S^r}_*$ categorifies $P^{S^r}:$
\[
\chi(\CH^{S^r}_*(K))=P^{S^r}(K).
\]
$\bullet$ \textbf{Anticommutativity:} The differentials $\{d^{S^r}_N\}$ anticommute\footnote{{\it cf.} comments following \eqref{frogtypes}}:
\[
d^{S^r}_Nd^{S^r}_M=-d^{S^r}_Md^{S^r}_N.
\]
$\bullet$ \textbf{Finite support:}
\[
\dim (\CH^{S^r}_*)< +\infty.
\]
$\bullet$ \textbf{Specializations:} For $N>1$, the homology of $\CH^{S^r}_*(K)$ with respect to $d^{S^r}_N$ is isomorphic to $\CH^{sl(N),S^r}(K)$:
\[
\left(\CH^{S^r}_*(K), d^{S^r}_N\right) \cong \CH^{sl(N),S^r}(K).
\]
$\bullet$ \textbf{Canceling differentials: } The differentials $d^{S^r}_1$ and $d^{S^r}_{-r}$  are \textit{canceling}:
the homology of $\CH^{S^r}_*(K)$ with respect to the differentials $d^{S^r}_1$ and $d^{S^r}_{-r}$ is one-dimensional,
with the gradings of the remaining generators being simple invariants of the knot $K$.\\

\noindent$\bullet$ \textbf{Vertical Colored differentials: } The differentials $d^{S^r}_{1-k}$, for $1 \le k \le r-1$, have $\a$-degree $-1$, and the homology of $\CH^{S^r}_*(K)$ with respect to the differential $d^{S^r}_{1-k}$ is isomorphic, after simple regrading that preserves $\a$- and $t$-gradings, to the $k$-colored homology $\CH^{S^{k}}_*(K)$.\\

\noindent$\bullet$ \textbf{Universal Colored differentials: } For any positive integer $m$, with $m<r$, the differentials $d_{r\to m}$
have $\a$-degree zero, and the homology of $\CH^{S^r}_*(K)$ with respect to the colored differential $d_{r\to m}$ is isomorphic (after regrading) to the $m$-colored homology $\CH^{S^m}_*(K)$:
\[
\left(\CH^{S^r}_*(K), d_{r\to m}\right) \cong \CH^{S^m}_*(K).
\]

\end{conjecture}
\vskip 1cm

A combinatorial definition of a triply-graded theory with the structure given in Conjecture \ref{con2},
as well as of the homologies $\CH^{sl(N),S^r}(K)$ for $r>1$ and $N>2$, still does not exist in the literature.

Even though there is no such combinatorial definition, one can use any combination of the above axioms as a definition,
and the remaining properties as consistency checks. In particular, one can obtain various consequence of the Conjecture \ref{con2}
and properties of the triply-graded homology $\CH^{S^r}$, along with the predictions for the triply-graded homology of simple knots.

In the rest of this section we give a summary of these properties, including some non-trivial checks.\\

\subsection{A word on grading conventions}
\label{sec:gradings}

So far we summarized the general structural properties of the colored knot homology.
Now we are about to make it concrete and derive explicit predictions for colored homology groups of simple knots.
This requires committing to specific grading conventions, as well as other choices that may affect the form of the answer.
It is important to realize, however, that none of these affect the very existence of the structural properties,
which are present with any choices and merely may look different.
While some of these choices will be discussed in section \ref{sec:comparison}, here we focus on

\begin{itemize}

\item
choices that associate various formulae to a Young tableaux $\lambda$ versus its transpose $\lambda^t$;

\item
choices of grading, {\it e.g.} grading conventions used in this paper (that we sometimes refer to as ``old'')
and grading conventions used in most of the existent literature \cite{DGR,AShakirov,DMMSS}
(that we sometimes call ``new'' in view of the forthcoming work \cite{inprogress} based on this choice).

\end{itemize}

The first choice here breaks the symmetry (``mirror symmetry'') between representations $S^r$ and $\Lambda^r$.
Indeed, since in view of the Conjecture \ref{con0}
the triply-graded homologies associated with these representations are essentially identical
and can be packaged in a single theory $\CH^r$, one has a choice whether $S^r$ homologies arise for $N>0$ or $N<0$.

The second choice listed here starts with different grading assignments, but turns out to be exactly the same as the first choice.
In other words, the ``old'' gradings and ``new'' gradings are related by ``mirror symmetry.''
Another way to describe this is to note, that in grading conventions of this paper the $S^r$-colored superpolynomials
are related (by a simple change of variables)
to the $\Lambda^r$-colored invariants that one would find by following
the same steps in grading conventions of {\it e.g.} \cite{DGR,AShakirov,DMMSS}:
\be
\CP^{S^r}_{\text{here}} \; = \; \CP^{\Lambda^r}_{\text{elsewhere}}
\ee
Note, that the $S^r$-colored invariant is related to the $\Lambda^r$-colored invariant, and vice versa.
The explicit change
of variables in this transformation is sensitive to even more elementary redefinitions,
such as $\a \to \a^2$ and $q \to q^2$ which is ubiquitous in knot theory literature.
For example, with one of the most popular choices of $\a$- and $q$-grading,
the transformation of variables / gradings looks like:
%$$
%\a \; \mapsto \; A \left( \frac{q}{t} \right)^{3/2} \,, \qquad
%q \; \mapsto \; \frac{1}{t} \,, \qquad
%t \; \mapsto \; - \sqrt{\frac{t}{q}} \,.
%$$
%\bea
%a & = & A \left( \frac{q_1}{q_2} \right)^{3/2} \,, \nonumber \\
%q & = & \frac{1}{q_2} \,, \label{GSvarchng} \\
%t & = & - \sqrt{\frac{q_2}{q_1}} \, \nonumber
%\eea
\bea
A & \mapsto & \a t^3 \,, \nonumber \\
q & \mapsto & \frac{1}{q t^2} \,, \label{GSvarchng} \\
t & \mapsto & \frac{1}{q} \,. \nonumber
\eea

The moral of the story is that, besides the grading conventions used in the earlier literature,
the present paper offers yet another choice of grading conventions consistent with all the structural properties.
And the relation between the two grading conventions can be viewed as a manifestation of mirror symmetry \eqref{mirsym}.
Keeping these words of caution in mind, now let us take a closer look at the structure
of the colored knot homology.

\subsection{Consequences of Conjecture \ref{con2}}\label{s42}

First of all, our main Conjecture \ref{con2} implies the Conjecture \ref{con1}.
Indeed, in order to be consistent with the specialization $\a=q^N$ from (\ref{coneq1}),
the $q$-degree of the differential $d^{S^r}_N$ must be proportional to $N$.
Since $\CH^{S^r}_*$ has finite support, this leads to the Conjecture \ref{con1},
with $\CP^{S^r}(K)$ being the Poincar\'e polynomial of $\CH^{S^r}_*(K)$.\\

More precisely, the differentials $d^{S^r}_N$, $N\ge 1$, are expected to have the following degrees:
\[
\deg (d^{S^r}_N) = (-1,N,-1),\quad N>0,
\]
which is consistent with the specialization $\a=q^N$ and the formula \eqref{WWqgrading}
that determines the $q$-grading of the corresponding differential in the doubly-graded theory.
In fact, the differential $d^{S^r}_N$ acts on the the following bi-graded chain complex:
\[
\CC^{sl(N),S^r}_{p,k}=\bigoplus_{iN+j=p} \CH^{S^r}_{i,j,k},
\]
and has $q$-degree 0, and $t$-degree $-1$.
The homology of $\CC^{sl(N),S^r}$ with respect to $d^{S^r}_N$ is isomorphic to $\CH^{sl(N),S^r}$.\\

In general, the degrees of the differentials $d^{S^r}_N$, for $N\in \Z$ are given by:

\begin{eqnarray*}
\deg(d^{S^r}_N)&=&(-1,N,-1),\quad N\ge 1-r,\\
\deg(d^{S^r}_N)&=&(-1,N,-3),\quad N \le -r.
\end{eqnarray*}
We note that for every $r\ge 1$, and every $N\in \Z$, the degree of the differential $d^{S^r}_N$ has the form $\deg(d^{S^r}_N)=(-1,N,\ast)$.

\subsubsection{Canceling differentials}

Canceling differentials appear in all conjectural triply-graded theories, including the ordinary HOMFLY homology and the Kauffman homology.
The defining property of a canceling differential is that the homology of the triply-graded theory with respect to this differential is ``trivial",
{\it i.e.} isomorphic to the homology of the unknot. In reduced theory, this means that the resulting homology is one-dimensional.
Furthermore, the degree of the remaining generator depends in a particularly simple way on the knot.\\

In the case of the colored HOMFLY homology $\CH^{S^r}$, the canceling differentials are $d^{S^r}_1$ and $d^{S^r}_{-r}$. Their degrees are:
\begin{eqnarray*}
\deg(d^{S^r}_1)&=&(-1,1,-1),\\
\deg(d^{S^r}_{-r})&=&(-1,-r,-3).
\end{eqnarray*}
Note that for $r=1$ this agrees with the gradings of the canceling differentials
in the ordinary triply-graded HOMFLY homology.
(Keep in mind, though, the conventions we are using in this paper, see Remark \ref{rem1}.)
For either of the two canceling differentials, the degree of the surviving generator depends only
on the $S$-invariant\footnote{Again, the value that we are using here is half of the value defined in \cite{DGR}.} of a knot $K$,
introduced in \cite{DGR}. In particular, the surviving generators have the following $(\a,q,t)$-degrees:
\begin{eqnarray}
\deg \left(\CH^{S^r}_*(K), d^{S^r}_1 \right) & = &(rS,-rS,0), \label{surviving} \\
\deg \left(\CH^{S^r}_*(K), d^{S^r}_{-r} \right) & = &(rS,r^2 S,2rS) \,. \nonumber
\end{eqnarray}
Note, that the remaining generator with respect to $d^{S^r}_1$ has $t$-degree equal to zero.\\

\subsubsection{Vertical Colored differentials}
Arguably, the most interesting feature of the colored triply-graded theory is the existence of colored differentials.
They allow to pass from the homology theory for a representation $R = S^r$ to the homology theory for another representation $R' = S^m$, with $m<r$.

The first group of colored differentials are ``vertical" colored differentials $d^{S^r}_{1-k}$, for $1\le k\le r-1$. As said before, the degrees of these differentials are
\be
\deg(d^{S^r}_{1-k})=(-1,1-k,-1),\quad 1\le k \le r-1.
\ee

The homology of $\CH^{S^r}$ with respect to the differential $d^{S^r}_{1-k}$, for any $k=1,\ldots,r-1$, is (up to a simple regrading) isomorphic to $\CH^{S^k}$.

In particular, up to an overall shift of the $\a$-grading, the $\a$- and $t$-gradings of the homologies match. More precisely, the Poincare polynomial of the homology $(\CH^{S^r},d^{S^r}_{1-k})$ satisfies:
\be
(\CH^{S^r},d^{S^r}_{1-k})(\a,q=1,t)=\a^{(r-k)S}\,\, \CP^{S^k} (\a,q=1,t),
\ee
where the $S$-invariant is defined by (\ref{surviving}).
The $q$-grading is controlled in the following way: the canceling differentials $d^{S^k}_1$ and $d^{S^k}_{-k}$ of $\CH^{S^k}$ correspond to two canceling differentials of $(\CH^{S^r},d^{S^r}_{1-k})$ of degrees $(-1,1+r-k,-1)$ and $(-1,-r,-3)$, respectively.

In the particular case of the differential $d^{S^r}_0$ that allows passage from $S^r$-colored homology to the uncolored $S^1$-homology, the explicit regrading is particularly simple:
\be
(\CH^{S^r},d^{S^r}_{0})(\a,q,t)=\a^{(r-1)S}\,\, \CP^{\tableau1} (\a,q^r,t).
\ee

\subsubsection{Universal Colored differentials}
The universal colored differentials in the triply-graded theory are ``universal" analogs
of the colored differentials in the doubly-graded theory \eqref{coloredHdH}:
\[
\CH^{sl(N),S^r} \; \leadsto \; \CH^{sl(N),S^{r-1}} \,.
\]
%that have been discussed in section \ref{sec:matrix}.
As explained in section \ref{sec:matrix}, these colored differentials come from the deformations of the Landau-Ginzburg potentials.\\

Since colored differentials are universal in the triply-graded theory --- {\it i.e.} they relate colored triply-graded homologies without even specializing to doubly-graded $sl(N)$ theories --- they should have zero $\a$-degree. This property distinguishes them clearly from the vertical colored differentials.

Furthermore, as explained in section \ref{sec:matrix},
the basic colored differential $d_{r\to (r-1)}$ should have $q$-degree equal to 1. Then, it is easy to see that consistency of the theory
also requires this differential to have zero $t$-grading. Combining all of these facts, we conclude that the $(\a,q,t)$-degree
of the differentials $d_{r\to (r-1)}$ is equal to $(0,1,0)$.\\

More generally, we expect that the degree of the differential $d_{r\to m}$ depends only on the difference $r-m$
and the homology of $\CH^{S^r}_*(K)$ with respect to the colored differential $d_{r\to m}$ is isomorphic (up to regrading) to $\CH^{S^m}_*(K)$.
The explicit form of the regrading for the colored differential $d_{2\to 1}$ is as follows: the Poincar\'e polynomial of the homology $(\CH^{\tableau2},d_{2\to 1})$ is equal to $\CP^{\tableau1}(\a^2,q^2,t^2q)$. Put differently,
\be
\left( \CH^{\tableau2} (K), d_{2\to 1} \right)_{2i,k+2j,2k} \; \cong \; \CH^{\tableau1}_{i,j,k} (K) \,.
\ee
In general, the explicit regrading is very subtle (unlike the case of the vertical colored differentials).

We point out that, for $m<m_1<r$,
the homology $((\CH^{S^r}, d_{r\to m_1}), d_{m_1\to m})$ does not need to coincide with the homology $(\CH^{S^r},d_{r\to m})$.

\subsection{Second symmetric representation}
\label{s2r}

In this subsection we focus on the case $r=2$, {\it i.e.} on the homology $\CH^{\tableau2}(K)$.
Specialization of the above mentioned properties to $r=2$ gives:

\begin{itemize}

\item
Two canceling differentials $d^{\tableau2}_1$ and $d^{\tableau2}_{-2}$ that have degrees:
\begin{eqnarray*}
\deg(d^{\tableau2}_1)&=&(-1,1,-1),\\
\deg(d^{\tableau2}_{-2})&=&(-1,-2,-3).
\end{eqnarray*}

\item
The generator that survives the differential $d^{\tableau2}_1$ has degree $(2S,-2S,0)$.

\item
The generator that survives the differential $d^{\tableau2}_{-2}$ has degree $(2S,4S,4S)$.

\item
The vertical colored differential $d^{\tableau2}_0$ has degree $(-1,0,-1)$.

\item
The Poincar\'e polynomial of the homology $(\CH^{\tableau2}(K),d^{\tableau2}_0)$ is equal to $\a^S\CP^{\tableau1}(K)(\a,q^2,t)$.

\item
The colored differential $d_{2\to 1}$ has degree $(0,1,0)$.

\item
The Poincar\'e polynomial of the homology $(\CH^{\tableau2}(K),d_{2\to 1})$ is equal to $\CP^{\tableau1}(K)(\a^2,q^2,t^2q)$.\\

\end{itemize}

In addition, the homology of $\CH^{\tableau2}(K)$ with respect to the differential $d_2^{\tableau2}$
should be isomorphic (after specialization $\a=q^2$) to $\CH^{sl(2),\tableau2}(K)$.
To find the latter homology one can use the isomorphism \eqref{sl2so3} with $\CH^{so(3),V} (K)$ which, in turn,
can be obtained from the triply-graded Kauffman homology $\CH^{\mathrm{Kauff}}(K)$ studied in \cite{GWalcher}.
Indeed, the doubly-graded homology $\CH^{so(3),V}(K)$ is isomorphic to the homology of $\CH^{\mathrm{Kauff}}(K)$
with respect to the corresponding differential $d_3$ from \cite{GWalcher}, after the specialization $\lambda=q^2$.
Usually, that differential $d_3$ acts trivially; in particular, this is the case for all knots that we analyze below.\\

The structure of the homology $\CH^{\tableau2}(K)$ with the above differentials allows us to compute it for various small knots,
as we shall illustrate next.\\

\subsubsection{$\CH^{S^2}$ and $\CP^{S^2}$ for small knots}

The homology $\CH^{\tableau2}(K)$ and the superpolynomial $\CP^{\tableau2}(K)(\a,q,t)$
(= the Poincar\'e polynomial of $\CH^{\tableau2}(K)$)
must satisfy the following properties:

\begin{itemize}

\item
specialization to $t=-1$ gives the reduced $\tableau2$-colored HOMFLY polynomial $P^{\tableau2}(K)$.\footnote{In order to find the colored HOMFLY polynomial $P^{\tableau2}(K)$ one can use {\it e.g.} equation (3.25) in \cite{LMV} and the values for the BPS invariants $\hat{N}_{\tableau2,g,Q}$ tabulated in section 6 of that paper. The result gives the unreduced two-variable colored HOMFLY polynomial. In order to find the reduced polynomial, one should divide the unreduced polynomial by:
\[
\frac{(\a -\a^{-1}) (\a q -\a^{-1}q^{-1}) }{(q-q^{-1})(q^2-q^{-2})} \,.
\]
The results from \cite{LMV} enable us to compute the reduced $\tableau2$-colored HOMFLY polynomial for the knots $3_1$, $4_1$, $5_1$ and $6_1$.
Another useful source of the colored HOMFLY polynomials and their specializations to $\a=q^2$ and $\a=q^3$ is the KnotAtlas \cite{KAt}, which the reader may want to consult for many other knots.}

\item
specialization to $\a=q^2$ gives the Poincar\'e polynomial $\CP^{\tableau2}_2(K)$ of the homology $\CH^{sl(2),\tableau2}(K)$. This homology is isomorphic to $\CH^{so(3),V}(K)$. To find the latter one, we use the results from Table 3 of \cite{GWalcher}, if available.\footnote{In all examples we have computed, the values $i+j$ for all nontrivial Kauffman homology groups $\CH^{\mathrm{Kauff}}_{i,j,k}(K)$ have the same parity. Thus the differential $d_3$ on the Kauffman homology, which is of degree $(-1,2,-1)$, is trivial. Consequently, the Poincar\'e polynomial $\CP^{so(3),V}(K)(q,t)$ is equal to the $\lambda=q^2$ specialization of the Poincar\'e superpolynomial of the Kauffman homology $\CP^{\mathrm{Kauff}}(K)(\lambda=q^2,q,t)$ in all our examples.}

\item
$\CH^{\tableau2}(K)$ comes equipped with the differentials described in section \ref{s2r}.

\end{itemize}

\noindent
These requirements are more than sufficient to determine the colored superpolynomial for many small knots.
As the first example, we consider the trefoil knot:
\begin{example}{The trefoil knot $3_1$}
The reduced $\tableau2$-colored HOMFLY polynomial of the trefoil knot is equal to (see {\it e.g.} \cite{LMV,LZ}):
\[
P^{\tableau2}(3_1) = \a^{2}q^{-2} + \a^2q + \a^2q^2 + \a^2q^4 - \a^3 - \a^3q - \a^3 q^3 - \a^3 q^4 + \a^4q^3 \,.
\]
The homology $\CH^{so(3),V}(3_1)$ is computed in \cite{GWalcher} (see eq. (6.14) and Table 3). Hence we have:
\[
\CP^{\tableau2}_2(3_1)=q^2+q^5t^2+q^6t^2+q^6t^3+q^7t^3+q^8t^4+q^9t^5+q^{10}t^5+q^{11}t^6
\]
{}From these two expressions we immediately deduce the colored superpolynomial of the trefoil\footnote{As discussed in
section \ref{sec:gradings}, there are two different possibilities for grading conventions.
Besides the grading conventions discussed in most of this paper, there are also ``new'' grading conventions
where the $\a$ and $q$ degrees are both twice the value of the corresponding degrees in the conventions that we are using in 
this paper, while the $t$-degree change is more subtle.
The value of the colored superpolynomial of the trefoil in the ``new'' gradings is given by
\[
\CP^{\tableau2}_{\text{new grad.}}(3_1)= \a^4(q^{-4} +q^2t^4+q^4t^6 +q^8t^8) + \a^6 (t^5+q^2t^7 + q^6t^9+q^8t^{11}) +\a^8q^6t^{12}.
\]
We note that in these gradings, the answer coincides with \cite{AShakirov,DMMSS}.}:
\begin{equation}
\CP^{\tableau2}(3_1)= \a^2(q^{-2} +qt^2+q^2t^2 +q^4t^4) + \a^3 (t^3+qt^3 + q^3t^5+q^4t^5) +\a^4q^3t^6.
\end{equation}
Note that its specializations to $t=-1$ and $\a=q^2$ give $P^{\tableau2}(3_1)$ and $\CP^{\tableau2}_2(3_1)$, respectively.
Moreover, the corresponding homology $\CH^{\tableau2}(3_1)$ also enjoys the action of two canceling differentials
and one colored differential.
In order to visualize this homology, we represent each generator by a dot in the $(\a,q)$-plane, with a label denoting its $t$-grading.
In Figure \ref{fig:colored31}, the canceling differential $d^{\tableau2}_1$ is represented by a blue arrow,
the canceling differential $d^{\tableau2}_{-2}$ is represented by a red arrow,
the colored differential $d_{2\to 1}$ is represented by a magenta arrow, while the vertical colored differential $d^{\tableau2}_0$
is represented by dashed light blue arrow.

\begin{figure}[htb]
\centering
\includegraphics[width=4.5in]{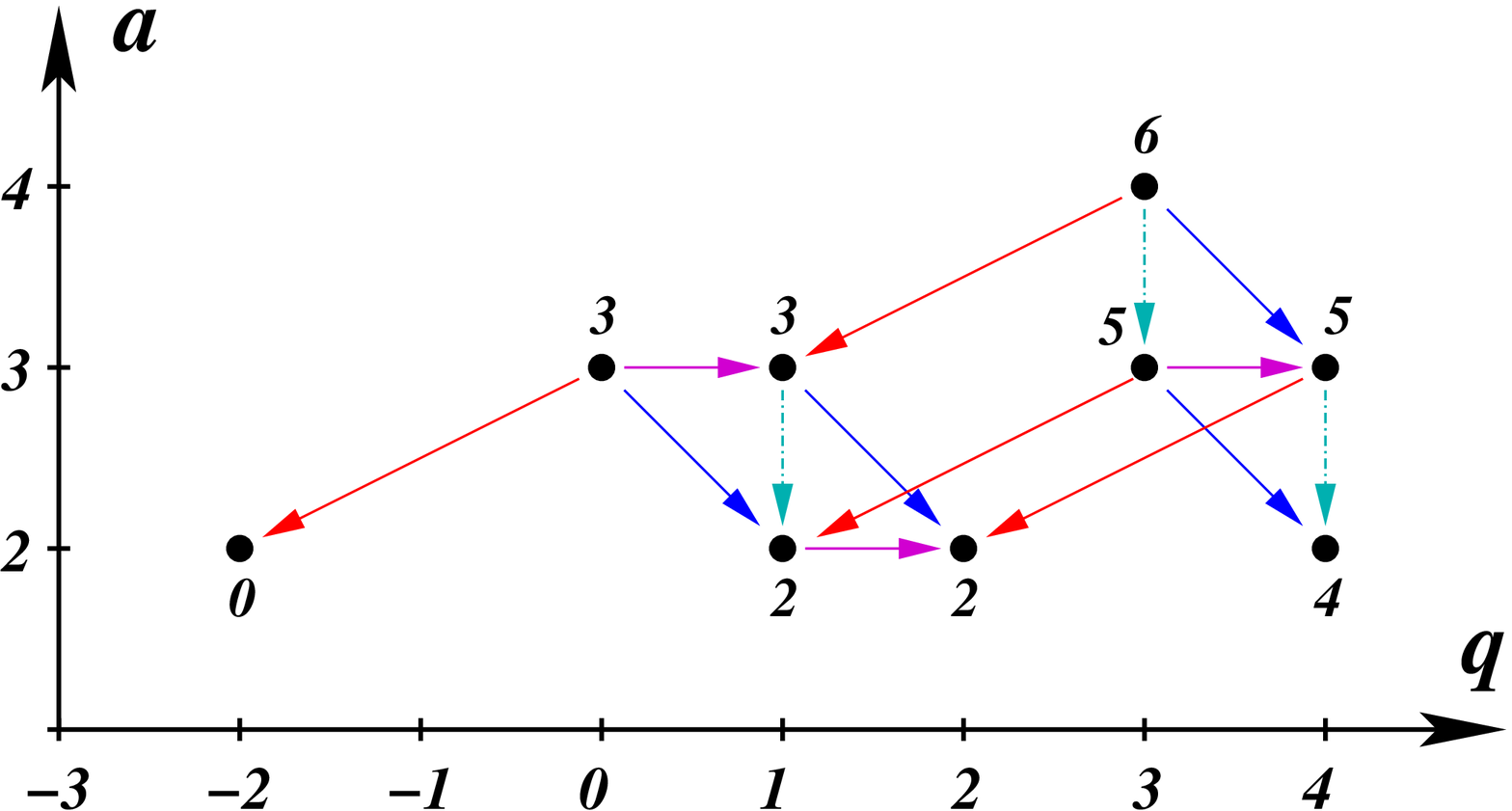}
\caption{The reduced $S^2$-colored homology of the trefoil knot.}
\label{fig:colored31}
\end{figure}

The generator that survives $d_1^{\tableau2}$ has degree $(2,-2,0)$, while the one that survives $d_{-2}^{\tableau2}$ has degree $(2,4,4)$.
Both are consistent with the $S$-invariant of the trefoil $S(3_1)=1$ and the general discussion in section~\ref{s2r}.
The Poincar\'e polynomial of the homology with respect to the colored differential $d_{2\to 1}$ is equal to:
\[
\a^2q^{-2} + \a^2q^4t^4 + \a^4q^3t^6 \,,
\]
while the Poincar\'e polynomial of the homology with respect to the vertical colored differential $d^{\tableau2}_{0}$ is equal to:
\[
\a^2 q^{-2} + \a^2 q^2 t^4 + \a^3 t^3 \,.
\]
A careful reader will notice that the last two expressions are equal to $\CP^{\tableau1}(3_1)(\a^2,q^2,t^2 q)$ and $\a \CP^{\tableau1}(3_1)(\a,q^2,t)$, respectively,
where $\CP^{\tableau1}(3_1)(\a,q,t)$ is the ordinary superpolynomial, whose explicit form is written in \eqref{p31}.
\end{example}

This computation can be easily extended to many other small knots.
We list the results for all prime knots with up to 6 crossings in Tables \ref{tablica} and \ref{tablica2}.
The fact that the structure described in this section works beautifully for all knots with up to 6 crossings is already
an impressive test of our main Conjecture \ref{con2}.
However, to convince even hard-boiled skeptics, in appendix \ref{sec:942} we carry out a much more challenging
computation of the colored HOMFLY homology for ``thick'' knots $8_{19}$ and $9_{42}$.

We notice that all computations of colored homologies in this paper are done by hand, only by using the existence and properties of the
differentials described in this section. Moreover, in majority of the cases only a few of the differentials  are used to obtain the result, which than matched perfectly all the remaining properties.  \\

\begin{table}\begin{center}
\begin{tabular}{|c|l|}
\hline
\rule{0pt}{5mm}
Knot & $\CP^{\tableau{2}}$ \\[3pt]
\hline
\hline
\rule{0pt}{5mm}
$3_1$ & $\a^2 (q^2 t^2 + q^{-2} + q t^2 + q^4 t^4) + \a^3 (q^4 t^5 + q^3 t^5 + t^3 + q t^3) + \a^4 q^3 t^6$ \\[3pt]
\hline
\rule{0pt}{5mm}
$4_1$ & $\a^{-2}q^{-2}t^{-4} + (\a^{-1}q^{-3}+\a^{-1}q^{-2})t^{-3} + (q^{-3}+\a^{-1}q^{-1}+\a^{-1})t^{-2}  +$\\[3pt]
\rule{0pt}{5mm}
& $+(q^{-2}+q^{-1}+\a^{-1}+\a^{-1}q)t^{-1} + (q^{-1}+3+q) + (q^2+q+\a+\a q^{-1})t +$\\[3pt]
\rule{0pt}{5mm}
& $+ (q^3 + \a q+\a)t^2 + (\a q^3+\a q^2)t^3 + \a^2q^2t^4$ \\[3pt]
\hline
\rule{0pt}{5mm}
$5_1$ & $\a^4q^{-4}+(\a^4q^{-1}+\a^4)t^2 + (\a^5q^{-2}+\a^5q^{-1})t^3 +$\\[3pt]
\rule{0pt}{5mm}
& $+(\a^4q^2+\a^4q^3+\a^4q^4)t^4 +(\a^5q+2\a^5q^2+\a^5q^3)t^5+$\\[3pt]
\rule{0pt}{5mm}
& $+(\a^4q^5+\a^4q^6+\a^6q)t^6 + (\a^5q^4+2\a^5q^5+\a^5q^6)t^7+$\\[3pt]
\rule{0pt}{5mm}
& $+(\a^4q^8+\a^6q^4+\a^6q^5)t^8+(\a^5q^7+\a^5q^8)t^9+\a^6q^7t^{10}.$ \\[3pt]
\hline
\rule{0pt}{5mm}
$5_2$ & $\a^2q^{-2}+(\a^2q^{-1}+\a^2)t+(2\a^2q+\a^2q^2+\a^3q^{-2}+\a^3q^{-1})t^2+$\\[3pt]
\rule{0pt}{5mm}
& $+(\a^2q^2+\a^2q^3+2\a^3+2\a^3q)t^3+(\a^2q^4+2\a^3q+3\a^3q^2+\a^3q^3+\a^4)t^4+$\\[3pt]
\rule{0pt}{5mm}
& $+(2\a^3q^3+2\a^3q^4+\a^4+2\a^4q+\a^4q^2)t^5+$\\[3pt]
\rule{0pt}{5mm}
& $+(\a^3q^4+\a^3q^5+\a^4q^2+3\a^4q^3+\a^4q^4)t^6+$\\[3pt]
\rule{0pt}{5mm}
& $+(\a^4q^3+2\a^4q^4+\a^4q^5+\a^5q^2+\a^5q^3)t^7+$\\[3pt]
\rule{0pt}{5mm}
& $+(\a^4q^6+\a^5q^3+\a^5q^4)t^8+(\a^5q^5+\a^5q^6)t^9+\a^6q^5t^{10}.$ \\[3pt]
\hline
\end{tabular}\end{center}
\caption{Colored superpolynomial for prime knots with up to 5 crossings.}
\label{tablica}
\end{table}

\begin{table}\begin{center}
\begin{tabular}{|c|l|}
\hline
\rule{0pt}{5mm}
Knot & $\CP^{\tableau{2}}$ \\[3pt]
\hline
\hline
\rule{0pt}{5mm}
$6_1$ &$\textstyle{ \a^{-2}q^{-2}t^{-4}+(\a^{-1}q^{-3}+\a^{-1}q^{-2})t^{-3}+(\a^{-1}q^{-2}+2\a^{-1}q^{-1}+q^{-3}+\a^{-1})t^{-2}+}$\\[3pt]
\rule{0pt}{5mm}
& $+(q^{-3}+\a^{-1}+3q^{-2}+2q^{-1}+\a^{-1}q)t^{-1}+(\a q^{-3}+2q^{-2}+\a q^{-2}+5+q)+$\\[3pt]
\rule{0pt}{5mm}
& $+(\a q^{-2}+1+3q+3\a q^{-1}+2q^2+2\a)t+$\\[3pt]
\rule{0pt}{5mm}
& $+(\a q^{-1}+4\a+4\a q+q^3+\a^2q^{-1}+\a q^2)t^2+$\\[3pt]
\rule{0pt}{5mm}
& $+(\a^2q^{-1}+\a q+2\a^2+3\a q^2+2\a q^3+\a^2q)t^3+$\\[3pt]
\rule{0pt}{5mm}
& $+(\a^2+2\a^2q+\a q^3+3\a^2q^2+\a q^4+\a^2q^3)t^4+$\\[3pt]
\rule{0pt}{5mm}
& $+(\a^2q^2+2\a^2q^3+\a^3q+\a^3q^2+\a^2q^4)t^5+$\\[3pt]
\rule{0pt}{5mm}
& $+(\a^3q^2+\a^3q^3+\a^2q^5)t^6+(\a^3q^4+\a^3q^5)t^7+\a^4q^4t^8.$ \\[3pt]
\hline
\rule{0pt}{5mm}
$6_2$ & $\a^4q^6t^8+ (\a^3q^6+\a^3q^7)t^7+(\a^4q^3+\a^2q^7)t^6+(2\a^3q^3+2\a^3q^4)t^5+$\\[3pt]
\rule{0pt}{2mm}
& $+(\a^4+\a^2q^3+3\a^2q^4+\a^2q^5)t^4+(2\a^3+2\a^3q+\a q^4+\a q^5)t^3+(3\a^2q+\a^2q^2)t^2+$\\[3pt]
\rule{0pt}{2mm}
& $+(\a^3q^{-3}+\a^3q^{-2}+\a q+\a q^2)t+(\a^2q^{-3}+3\a^2q^{-2}+\a^{2}q^{-1}+q^2)+$\\[3pt]
\rule{0pt}{3mm}
& $+(\a q^{-2}+\a q^{-1})t^{-1}+ \a^2q^{-5}t^{-2}+(\a q^{-5}+\a q^{-4})t^{-3}+q^{-4}t^{-4}+$\\[3pt]
\rule{0pt}{2mm}
%& $+ \a^2q^{-5}t^{-2}+(\a q^{-5}+\a q^{-4})t^{-3}+q^{-4}t^{-4}+$\\[3pt]
%\rule{0pt}{3mm}
& $+(1+q)(1+\a^{-1}qt^{-1})(1+\a^{-1}q^{-2}t^{-3}) \times [\a^4q^4t^7+(\a^4q^3+\a^3q^4)t^6+$\\[3pt]
\rule{0pt}{3mm}
& $+(\a^4q+\a^3q^3)t^5+(\a^3q^2+\a^3q)t^4+\a^3t^3+(\a^3q^{-2}+\a^2)t^2].$\\[3pt]
\hline
\rule{0pt}{5mm}
$6_3$ & $ \a^2q^5t^6+(\a q^5+\a q^6)t^5+(\a^2q^2+q^6)t^4+(2\a q^2+2\a q^3)t^3+$\\[3pt]
\rule{0pt}{3mm}
& $+(\a^2q^{-1}+q+3q^2+q^3)t^2+(2\a q^{-1}+2\a+\a^{-1}q^3+\a^{-1}q^4)t$\\[3pt]
\rule{0pt}{3mm}
&$+(q^{-1}+5+q)+(\a q^{-4}+\a q^{-3}+2\a^{-1}+2\a^{-1}q)t^{-1}+(q^{-3}+3q^{-2}+q^{-1}+\a^{-2}q)t^{-2}+$\\[3pt]
\rule{0pt}{3mm}
&$+(2\a^{-1}q^{-3}+2\a^{-1}q^{-2})t^{-3}+(q^{-6}+\a^{-2}q^{-2})t^{-4}+(\a^{-1}q^{-6}+\a^{-1}q^{-5})t^{-5}+\a^{-2}q^{-5}t^{-6}+$\\[3pt]
\rule{0pt}{3mm}
&$+ (1+q)(1+\a^{-1}qt^{-1})(1+\a^{-1}q^{-2}t^{-3}) \times[\a^2q^3t^5+(\a^2q+\a q^3)t^4+   $\\[3pt]
\rule{0pt}{3mm}
& $+ (\a^2+\a q^2+\a q)t^3+3\a t^2+(\a q^{-1}+\a q^{-2}+1)t+(\a q^{-3}+q^{-1}) +q^{-3}t^{-1}].$\\[3pt]
\hline
\end{tabular}\end{center}
\caption{Colored superpolynomial for prime knots with 6 crossings.}
\label{tablica2}
\end{table}

\subsection{Differentials for higher symmetric representations}

Now let us consider the triply graded homology $\CH^{S^r}$ of knots and links colored by the representation $R = S^r$ with more general $r \ge 1$.
Much as in the case $r=2$ considered in the previous subsection, we expect that $\CH^{S^r}$ comes equipped with the following differentials:

\begin{itemize}

\item
canceling differential $d^{S^r}_1$ of  degree $(-1,1,-1)$, whose homology is one-dimensional and consists of a degree $(rS,-rS,0)$ generator;

\item
canceling differential $d^{S^r}_{-r}$ of degree $(-1,-r,-3)$, which leaves behind a one-dimensional homology with a generator of degree $(rS,r^2S,2rS)$;

\item
for every $1 \le k <r$, there exists a vertical colored differential $d^{S^r}_{1-k}$ of degree $(-1,1-k,-1)$, such that the homology of $\CH^{S^r}$ with respect to $d^{S^r}_{1-k}$ is isomorphic to $\CH^{S^k}$;

\item
for every $1 \le m < r$, there exists a universal colored differential $d_{r \to m}$ which, when acting on $\CH^{S^r}$, leaves behind the homology $\CH^{S^{m}}$. In particular, the colored differential $d_{r\to (r-1)}$ has degree $(0,1,0)$.

\end{itemize}

\subsubsection{Colored superpolynomial $\CP^{S^3}$ for the trefoil}
It can be computed by requiring that its specialization to $t=-1$ equals the reduced $S^3$-colored HOMFLY polynomial and that it enjoys the action of the canceling and the first colored differentials of appropriate degrees. In particular, according to the general rules \eqref{surviving}, we require that the remaining generator with respect to the $d^{\tableau3}_1$ action has degree $(3,-3,0)$, while the remaining generator with respect to the action of $d^{\tableau3}_{-3}$ has degree $(3,9,6)$. For the colored differential $d_{3\to 2}$ we require that the remaining homology should have rank 9, just like $\CH^{\tableau2}(3_1)$. From these, we obtain the following result:
\begin{eqnarray*}
\CP^{S^3}(3_1)&=&\a^3q^{-3}+(\a^3q+\a^3q^2+\a^3q^3)t^2+(\a^4+\a^4q+\a^4q^2)t^3+\\
&&+(\a^3q^5+\a^3q^6+\a^3q^7)t^4+(\a^4q^4+2\a^4q^5+2\a^4q^6+\a^4q^7)t^5+\\
&&+(\a^5q^4+\a^5q^5+\a^5q^6+\a^3q^9)t^6+(\a^4q^8+\a^4q^9+\a^4q^{10})t^7+\\
&&+(\a^5q^8+\a^5q^9+\a^5q^{10})t^8+\a^6q^9t^9.
\end{eqnarray*}
Note, there exists a differential $d_{3\to 1}$ on $\CH^{S^3}(3_1)$ of degree $(0,4,2)$,
such that the homology with respect to this differential is of rank 3, as $\CH^{\tableau1}(3_1)$.

Also, there exists a differential $d^{\tableau3}_0$ of degree $(-1,0,-1)$ such that the Poincare polynomial of $(\CH^{\tableau3}(3_1),d^{\tableau3}_0)$ is equal to $\a^2\,\,\CP^{\tableau1}(3_1) (\a,q^3,t)$.

Finally, there exists a differential $d^{\tableau3}_{-1}$ of degree $(-1,-1,-1)$ such that the homology$(\CH^{\tableau3}(3_1),d^{\tableau3}_{-1})$ is isomorphic to $\CH^{\tableau2}(3_1)$.\\

In Appendix \ref{apb41} we compute also the $S^3$-colored homology of the figure-eight knot $4_1$.

\subsubsection{Size of the homology}
Computations show that for a knot $K$, the rank of the homology $\CH^{S^r}$ grows exponentially with $r$.
In particular, this makes the computation of the homologies $\CH^{S^r}(K)$ difficult for large $r$.
(In fact, even for $r>2$ the size of the homology is too big to make computations practical.)
To be more precise, for all thin and torus knots studied here we find:
\be
\rank \CH^{S^r} (K) \; = \; \left( \rank \CH^S (K) \right)^r \,.
\ee

%%%%%%%%%%%%%%%%%%%%%%%%%%%%%%%%%%%%%%%%%%%%%%%%%%%%%%%%%%%%%%%%%%%%%
\section{Mirror symmetry for knots}
\label{sec:mirror}

In this section, we observe a remarkable ``mirror symmetry'' relation \eqref{mirwild}
between two completely different triply-graded homology theories associated with symmetric and anti-symmetric representations of $sl(N)$,
which will allow us to formulate even a bigger theory that will contain both.
As a first step, however, we need to extend the discussion in section \ref{sec:reduced} to the HOMFLY homology colored
by anti-symmetric representations of $sl(N)$.

\subsection{Anti-symmetric representations}

Much as for the symmetric representation $S^r$, we can repeat the analysis for the anti-symmetric representations $\Lambda^r$ of $sl(N)$.\\

In particular, for every positive integer $r$ there exists a triply-graded homology theory $\CH^{\Lambda^r}(K)$, together with the collection of differentials $\{d^{\Lambda^r}_{N}\}$, $N\in\Z$, such that the homology with respect to $d^{\Lambda^r}_{N}$ is isomorphic to $\CH^{sl(N),\Lambda^r}(K)$. Moreover, it comes equipped with the collection of ``universal" colored differentials, like in the case of the symmetric representations. The homologies $\CH^{\Lambda^r}(K)$, together with all the differentials, satisfy the same properties as $\CH^{S^r}(K)$ from Conjecture \ref{con2}. \\

Again, we have two canceling differentials, this time $d^{\Lambda^r}_{-1}$
and $d^{\Lambda^r}_r$ of $(\a,q,t)$-degrees $(-1,-1,-3)$ and $(-1,r,-1)$, respectively.
The origin of these differentials is clear, and can be inferred either from deformations of B-model potentials, as in section \ref{sec:matrix},
or from basic representation theory. For example, the fact the representation $\Lambda^r$ of $sl(r)$ is trivial
gives rise to a canceling differential $d^{\Lambda^r}_r$.

Another basic fact is $\Lambda^r \cong \Lambda$ for $\g=sl(r+1)$, which leads to the relation
\be
\CH^{sl(r+1), \Lambda^r} (K) \; \cong \; \CH^{sl(r+1), \tableau{1}} (K) \,.
\ee
For the triply-graded theory $\CH^{\Lambda^r}(K)$, this relation implies that
the $\a=q^{r+1}$ specializations of the homologies $(\CH^{\Lambda^r}(K), d^{\Lambda^r}_{r+1})$
and $(\CH^{\tableau1}(K),d^{\tableau1}_{r+1})$ should be isomorphic.\\

Like in the case of symmetric representations, all the required properties allow computation
of the anti-symmetric homology for various small knots. Below, we provide the details for the trefoil knot.

Using the isomorphism $so(6) \cong sl(4)$ under which the vector representation of $so(6)$
is identified with the anti-symmetric representation of $sl(4)$, we conclude
\be
\CH^{sl(4), \tableau{1 1}} (K) \; \cong \; \CH^{so(6),V} (K)
\ee
{}From this relation\footnote{Here, in comparing the two homology theories we take into account that $q$-gradings differ by a factor of~2.} we immediately find
\be
\CH^{sl(4), \tableau{1 1}} (3_1) =
q^4 + q^{6} t^2 + q^{7} t^2 + q^{8} t^3 + q^{9} t^3 + q^{10} t^4 + q^{11} t^5 + q^{12} t^5 + q^{13} t^6
\ee

Also, for $\frak g = sl(3)$ we have $\Lambda^2 \cong \Lambda$, which implies another useful relation
\be
\CH^{sl(3), \tableau{1 1}} (K) \; \cong \; \CH^{sl(3), \tableau{1}} (K)
\ee
For the trefoil, this gives:
\be
\CH^{sl(3), \tableau{1 1}} (3_1) = q^2 + q^4 t^2 + q^{6} t^3
\label{sl3anti}
\ee
Combining this data with the colored HOMFLY polynomial
\be
P^{\tableau{1 1}} (\a,q) \; = \;
\a^2 (q^{-4} + q^{-2} + q^{-1} + q^2)
- \a^3 (q^{-4} + q^{-3} + q^{-1} + 1)
+ \a^4 q^{-3}
\ee
we easily find the anti-symmetric version of the superpolynomial for the trefoil knot:
\be
\CP^{\tableau{1 1}} (3_1) \; = \;
\a^2 (q^{-4} + q^{-2} t^2 + q^{-1} t^2 + q^2 t^4)
+ \a^3 (q^{-4} t^3 + q^{-3} t^3 + q^{-1} t^5 + t^5)
+ \a^4 q^{-3} t^6
\ee

\begin{figure}[htb]
\centering
\includegraphics[width=4.5in]{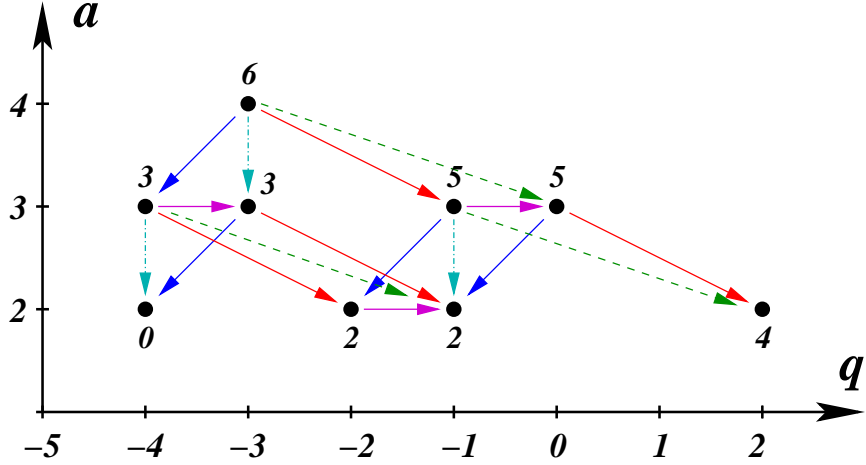}
\caption{The reduced $\Lambda^2$-colored homology of the trefoil}
\label{fig:colored31A2}
\end{figure}

The homology $\CH^{\tableau{1 1}} (3_1)$ is shown on the Figure \ref{fig:colored31A2} below. It has the following differentials:

\begin{itemize}

\item canceling differential $d^{\tableau{1 1}}_{-1}$ of $(\a,q,t)$-degree $(-1,-1,-3)$

\item canceling differential $d^{\tableau{1 1}}_2$ of $(\a,q,t)$-degree $(-1,2,-1)$ reflects the fact that $\Lambda^2$ is trivial in $sl(2)$ theory

\item differential $d^{\tableau{1 1}}_3$ of $(\a,q,t)$-degree $(-1,3,-1)$ reflects the fact that $\Lambda^2 \cong \Lambda$ in $sl(3)$ theory
and gives, {\it cf.} \eqref{sl3anti}:
\be
\a^2 q^{-4} t^0 + \a^2 q^{-2} t^2 +\a^3 q^{-3} t^3
\; \overset{\underset{\a=q^3}{}}{=} \;
\CH^{sl(3), \tableau{1 1}} (3_1)
%q^2 + q^4 t^2 + q^6 t^3
\ee

\item differential $d^{\tableau{1 1}}_0$ of $(\a,q,t)$-degree $(-1,0,-3)$ gives:
\be
\a^2 q^{-2} t^2 + \a^2 q^2 t^4 + \a^3 q^0 t^5 = \a t^2 \CP^{\tableau1}(3_1)(\a,q^2,t).
\ee

\item universal differential $d^{\tableau{1 1}}_{2 \rightarrow 1}$ of $(\a,q,t)$-degree $(0,1,0)$ gives:
\be
\a^2 q^{-4} t^0 + \a^2 q^2 t^4 + \a^4 q^{-3} t^6 \; = \; \CP^{\tableau{1}} (\a^2,q^4,q^{-1} t^2)
\ee

\end{itemize}

\subsection{Mirror symmetry for knot homology}

By computing the triply-graded homologies $\CH^{S^r}(K)$ and $\CH^{\Lambda^r}(K)$ for various small knots,
we discover the following remarkable symmetry between these two classes of theories, labeled by $R = S^r$ and $R = \Lambda^r$:
\be\label{mirsym2}
\CH^{\Lambda^r}_{i,j,*}(K) \cong \CH^{S^r}_{i,-j,*}(K).
\ee
Furthermore, this symmetry extends to the differentials as well. More precisely, let
\be
\phi:\CH^{S^r}(K)\to \CH^{\Lambda^r}(K),
\ee
be the isomorphisms  from (\ref{mirsym2}). Then
\be\label{difsym}
\phi\, d^{S^r}_N = d^{\Lambda^r}_{-N}\, \phi,\quad N\in\Z.
\ee

As the first illustration of the mirror symmetry, let us compare the second symmetric and anti-symmetric homology for the trefoil knot.
{}From Figures \ref{fig:colored31} and \ref{fig:colored31A2} it is clear that ``mirror symmetry" is manifest both
for the homology (\ref{mirsym2}) and for the differentials (\ref{difsym}).
The explicit $t$-grading change in (\ref{mirsym2}) in this case is given~by:
\be\label{mirsym231}
\CH^{\tableau{1 1}}_{i,j,k}(3_1) \cong \CH^{\tableau2}_{i,-j,4i-k-4}(3_1)
\ee
for the trefoil knot.\\

The first implication of the mirror symmetry is that one can combine $\CH^{S^r}(K)$ and $\CH^{\Lambda^r}(K)$ into a single homology theory.
By setting $\CH^r(K)$ to be $\CH^{S^r}(K)$, we obtain the Conjecture \ref{con0}. More precisely, we conjecture the following:

\begin{conjecture}\label{mirrorcon}
For every positive integer $r$ there exists a triply-graded homology theory $\CH^{r}_*(K)=\CH^{S^r}_{i,j,k}(K)$,
that comes with a family of differentials $\{d^{r}_N\}$, with $N\in\Z$, and also with an additional collection of \textit{universal colored} differentials $d_{r\to m}$, for every $1\le m<r$, satisfying the following properties:\\

\noindent$\bullet$ \textbf{Mirror Symmetry}
\[
\CH^{r}_{i,j,*}(K)\cong\CH^{S^r}_{i,j,*}(K)\cong\CH^{\Lambda^r}_{i,-j,*}(K).
\]
\noindent$\bullet$ \textbf{Categorification: } $\CH^{r}_*$ categorifies $P^{S^r}$ and $P^{\Lambda^r}:$
\[
\chi(\CH^{r}_*(K))=P^{S^r}(K)(a,q)=P^{\Lambda^r}(K)(a,q^{-1}).
\]
$\bullet$ \textbf{Anticommutativity:} The differentials $\{d^{S^r}_N\}$ anticommute\footnote{{\it cf.} comments following \eqref{frogtypes}}:
\[
d^{r}_Nd^{r}_M=-d^{r}_Md^{r}_N.
\]
$\bullet$ \textbf{Finite support:}
\[
\dim (\CH^{r}_*)< +\infty.
\]
$\bullet$ \textbf{Specializations:} For $N>1$, the homology of $\CH^{r}_*(K)$ with respect to $d^{r}_N$ is isomorphic to $\CH^{sl(N),S^r}(K)$:
\[
\left(\CH^{r}_*(K), d^{r}_N\right) \cong \CH^{sl(N),S^r}(K).
\]
For $N \le -2r$, the homology of $\CH^{r}_*(K)$ with respect to $d^{r}_N$ is isomorphic to $\CH^{sl(-N),\Lambda^r}(K)$:
\[
\left(\CH^{r}_*(K), d^{r}_N\right) \cong \CH^{sl(-N),\Lambda^r}(K).
\]
$\bullet$ \textbf{Canceling differentials: } The differentials $d^{r}_1$ and $d^{r}_{-r}$  are \textit{canceling}:
the homology of $\CH^{r}_*(K)$ with respect to the differentials $d^{r}_1$ and $d^{r}_{-r}$ is one-dimensional. This
reflects the fact that $S^r$ representation of $sl(1)$ and $\Lambda^r$ representation of $sl(r)$ are trivial.\\

\noindent$\bullet$ \textbf{$sl(N)$ Colored differentials: } For every $1\le k \le r-1$, the homology of  $\CH^{r}_*(K)$ with respect to the differential $d^{r}_{-r-k}$ is isomorphic to $\CH^{k}(K)$:
\[
\left(\CH^{r}_*(K), d^{r}_{-r-k}\right) \cong \CH^{k}(K),\quad 1\le k\le r-1,
\]
reflecting the fact that $\Lambda^{r}\cong \Lambda^k$ for $sl(r+k)$, where $1\le k\le r-1$.\\

\noindent$\bullet$ \textbf{Vertical Colored differentials: } The differentials $d^{r}_{1-k}$, for $1 \le k \le r-1$, have $\a$-degree $-1$, and the homology of $\CH^{r}_*(K)$ with respect to the differential $d^{r}_{1-k}$ is isomorphic (after simple regrading that preserves $\a$- and $t$-gradings), to $\CH^{{k}}_*(K)$:
\[
\left(\CH^{r}_*(K), d^{r}_{1-k}\right) \cong \CH^{k}(K),\quad 1\le k\le r-1.
\]

\noindent$\bullet$ \textbf{Universal Colored differentials: } For any positive integer $m$, with $m<r$, the differentials $d_{r\to m}$have $\a$-degree zero, and the homology of $\CH^{r}_*(K)$ with respect to the colored differential $d_{r\to m}$ is isomorphic (after regrading) to $\CH^{m}_*(K)$:
\[
\left(\CH^{r}_*(K), d_{r\to m}\right) \cong \CH^{m}_*(K).
\]

\end{conjecture}

\begin{example}{$\CH^3$ for the trefoil knot}
As another example of the symmetry \eqref{mirsym2}, and this time for degree higher than $2$, let us consider $\CH^3(3_1)$.
Besides the two canceling differentials $d^3_1$ and $d^3_{-3}$, we also have the differential $d^3_{-4}$ of degree $(-1,-4,-3)$.
The homology with respect to this differential is equal to
\be\label{form331}
(\CH^3(3_1),d^3_{-4}) = \a^3q^7t^4+\a^3q^9t^6+\a^4q^8t^7.
\ee
The explicit $t$-grading change in (\ref{mirsym2}) for the trefoil and $r=3$ is given by:
\be\label{mirsym331}
\CH^{\tableau{1 1 1}}_{i,j,k}(3_1) \cong \CH^{\tableau3}_{i,-j,4i-k-6}(3_1).
\ee
With this change of gradings, the ``mirror image" of (\ref{form331}) is equal to $\a^3q^{-9}t^0+\a^3q^{-7}t^{2}+\a^4q^{-8}t^3$ which,
according to the Conjecture \ref{con0}, should be equal to the homology $(\CH^{\tableau{1 1 1}}(3_1),d^{\tableau{1 1 1}}_{4})$.
In particular, for $\a=q^4$ it implies that the Poincar\'e polynomial of
$(\CH^{\tableau{1 1 1}}(3_1),d^{\tableau{1 1 1}}_{4})$ is equal to $q^3 + q^5 t^2 + q^8 t^3$.
The latter, in turn, is equal to $\CP^{\tableau1}(3_1)(\a=q^4,q,t)$ (see (\ref{p31})),
in agreement with the isomorphism $\Lambda^3\cong \Lambda$ for $sl(4)$.
\end{example}

In Appendix \ref{apb41} we compute $\CH^3$ of the figure-eight knot and show that it satisfies all of the properties from conjecture \ref{mirrorcon}.

It is straightforward to check that the triply-graded colored HOMFLY homologies of other knots
that we computed in Section \ref{sec:reduced} satisfy all the required properties of $\CH^2$.\\

As for the $t$-grading change in (\ref{mirsym2}) for an arbitrary knot,
$\phi$ sends a generator $x$ of $(a,q,t)$-degree $(i,j,k)$ to a generator of degree $(i,-j,4i-k+2r\delta'(x))$.
Here $\delta'(x)$ is a certain grading of the generator $x$, generalizing the $\delta$-grading of the ordinary HOMFLY homology $\CH^{\tableau1}$.
In the case of thin knots, $\delta'$-grading of all generators is equal to the $S$-invariant of knots.\\

Now, since the $t$-gradings of $x$ and $\phi(x)$ have the same parity, by decategorifying (\ref{mirsym2})
we get the following simple and beautiful relation between the colored HOMFLY polynomials:
\be\label{mirsymp}
P^{S^r}(K)(\a,q)=P^{\Lambda^r}(K)(\a,q^{-1}).
\ee
To the best of our knowledge, this relation has not been observed before.
It generalizes the symmetry $q \leftrightarrow q^{-1}$ of the (ordinary) HOMFLY polynomial.\\

Based on the above observations about the mirror symmetry for symmetric and anti-symmetric representations,
we speculate that this symmetry extends to arbitrary representations: for a representation $R$ of $sl(N)$
that corresponds to a partition $\lambda$, we conjecture:
\be\label{mirgeneral}
\CH^{\lambda} (K) \; {\cong}\; \CH^{\lambda^t} (K) \,,
\ee
where $\lambda^t$ is the dual (transpose) partition of $\lambda$.
Here, we also tacitly assume that for every partition $\lambda$ there exists
a triply-graded homology theory $\CH^{\lambda}(K)$ categorifying the $\lambda$-colored HOMFLY polynomial.

Furthermore, decategorifying the isomorphism (\ref{mirgeneral}) we obtain
the following symmetry of the colored HOMFLY polynomials (see appendix \ref{sec:notation} for conventions):
\be\label{mirsympgen}
P^{\lambda}(K)(\a,q) \; = \; P^{\lambda^t}(K)(\a,q^{-1}) \,.
\ee

%%%%%%%%%%%%%%%%%%%%%%%%%%%%%%%%%%%%%%%%%%%%%%%%%%%%%%%%%%%%%%%%%%%%%%%%%%%%%%%%%%%%%
\subsection{Physical interpretation}
\label{sec:physmir}

The only evidence for the mirror symmetry \eqref{mirgeneral} and for its polynomial version \eqref{mirsympgen}
comes from the physical interpretation of knot homologies / polynomials in terms of BPS invariants.
Indeed, the symmetry \eqref{mirsympgen} of the reduced colored HOMFLY polynomial is a direct consequence
of the corresponding symmetry for the unreduced colored HOMFLY polynomial:
\be
\bar P^{\lambda^t}(K)(\a,q) \; = \; (-1)^{|\lambda|} \; \bar P^{\lambda}(K)(\a,q^{-1}) \,,
\label{mirunnorm}
\ee
where $|\lambda|$ is the total number of boxes in the partition $\lambda$.
This symmetry, in turn, follows\footnote{Another way to derive \eqref{mirunnorm}
is to use the properties of the Clebsch-Gordon coefficients for the symmetric group
$C_{\lambda \mu^t \nu^t} = C_{\lambda \mu \nu}$
and the characters $S_{\lambda^t} (q) = (-1)^{|\lambda|-1} S_{\lambda} (q^{-1})$ in eq.(2.6) of \cite{MV},
which describes the geometric origin of the $q$-dependence
in the colored HOMFLY polynomail.} from the explicit form of the colored HOMFLY polynomial \cite{OoguriV}:
\be
\bar P^{\lambda}(K)(\a,q) \; = \; \sum_{i,j} \frac{\a^i q^j}{q-q^{-1}} N^{\lambda}_{i,j} (K)
\label{HOMFLYBPS}
\ee
written in terms of the ordinary (that is, unrefined) BPS invariants $N^{\lambda}_{i,j} (K)$
and from the property of the integer BPS invariants \cite[eq. (2.17)]{LMV}:
\be
N^{\lambda^t}_{i,j} (K) \; = \; (-1)^{|\lambda|+1} \; N^{\lambda}_{i,-j} (K) \,.
\label{BPSmir}
\ee
Indeed, when combined, \eqref{HOMFLYBPS} and \eqref{BPSmir} imply that
under $q \leftrightarrow q^{-1}$ the unreduced colored HOMFLY polynomial of every knot has parity $(-1)^{|\lambda|}$.
In particular, this is true for the unknot. Hence, the normalized colored HOMFLY polynomial,
defined as the ratio of $\bar P^{\lambda}(K)$ and $\bar P^{\lambda}(\unknot)$,
enjoys the symmetry \eqref{mirsympgen}.

There is a refined / homological version of \eqref{mirunnorm} and \eqref{BPSmir} that leads to \eqref{mirgeneral}.
Much as the colored HOMFLY polynomial can be written in terms of the unrefined BPS invariants $N^{\lambda}_{i,j} (K)$,
the unnormalized superpolynomial of every knot $K$ can be expressed in terms of the refined integer BPS invariants \cite{GSV}:
\be
\bar \CP^{\lambda}(K)(\a,q,t) \; = \; \sum_{i,j,k} \frac{\a^i q^j t^k}{q-q^{-1}} D^{\lambda}_{i,j,k} (K) \,.
\label{superBPS}
\ee
The refined BPS invariants $D^{\lambda}_{i,j,k} (K)$ enjoy a symmetry that generalizes \eqref{BPSmir},
\be
D^{\lambda^t}_{i,j,*} (K) \; = \; D^{\lambda}_{i,-j,*} (K) \,,
\label{refBPSmir}
\ee
and follows from the CPT symmetry of the five-brane theory in \eqref{theoryA} or \eqref{theoryB}.
The normalized / reduced version of the symmetry \eqref{refBPSmir} is precisely \eqref{mirgeneral}.

Further details and interpretation of the mirror symmetry \eqref{mirgeneral} for the triply-graded
knot homology will appear elsewhere.

%%%%%%%%%%%%%%%%%%%%%%%%%%%%%%%%%%%%%%%%%%%%%%%%%%%%%%%%%%%%%%%%%%%%%

\section{Unreduced colored HOMFLY homology}
\label{sec:unreduced}

Here we compute the unreduced colored superpolynomial and the colored HOMFLY polynomial
of the unknot and the Hopf link by using the refined topological vertex approach from \cite{GIKV}.
The formulas obtained there are partition functions, presented in the form of the quotient of two infinite series.
Below we find the explicit closed form expressions for the unreduced $S^r$-colored HOMFLY homology of the unknot and the Hopf link.
More precisely, we evaluate eq. (67) of \cite{GIKV}, according to which the unreduced superpolynomial
(= the Poincar\'e polynomial of the unreduced triply-graded colored homology)
of the Hopf link with components colored by partitions $\lambda$ and $\mu$ is given by:
\begin{equation}
\label{eq67}
\bar{\mathcal{P}}^{\lambda\mu}(\text{Hopf})=(-1)^{|\lambda|+|\mu|} \left(\frac{q_1}{q_2}\right)^{|\lambda||\mu|} \left(Q^{-1}\sqrt{\frac{q_1}{q_2}}\right)^{\frac{|\lambda|+|\mu|}{2}}\times\frac{Z_{\lambda\mu}}{Z_{\emptyset\emptyset}} \,,
\end{equation}
where
\[
Z_{\lambda\mu}=\sum_{\nu} (-Q)^{|\nu|} q_2^{\frac{||\nu||}{2}} q_1^{\frac{||\nu^t||}{2}} \tilde{Z}_{\nu}(q_1,q_2)\tilde{Z}_{\nu^t}(q_2,q_1) s_{\lambda}(q_2^{-\rho}q_1^{-\nu^t}) s_{\mu}(q_2^{-\rho}q_1^{-\nu^t}) \,.
\]
The unreduced superpolynomial of the unknot colored by $\lambda$ is obtained by setting $\mu=\emptyset$ in (\ref{eq67}).

The change of variables from topological strings variables $(Q,q_1,q_2)$ to knot theory variables $(\a,q,t)$ used in this paper is given by:
\begin{eqnarray}
\sqrt{q_2}&=&q, \\ \nonumber
\sqrt{q_1}&=&-tq, \\ \label{transf}
Q&=&-t\a^{-2}. \nonumber
\end{eqnarray}
In particular, the specialization $q_1=q_2$ corresponds to the specialization $t=-1$ in the homological knot invariants.

By expanding the product of the Schur functions as
\[
s_{\lambda}s_{\mu}=\sum_{\varphi} c_{\lambda,\mu}^{\varphi} s_{\varphi} \,,
\]
where $c_{\lambda,\mu}^{\varphi}$ is the Littlewood-Richardson coefficient, we obtain:
\[
Z_{\lambda\mu}=\sum_{\varphi} c_{\lambda,\mu}^{\varphi}  \sum_{\nu} (-Q)^{|\nu|} q_2^{\frac{||\nu||}{2}} q_1^{\frac{||\nu^t||}{2}} \tilde{Z}_{\nu}(q_1,q_2)\tilde{Z}_{\nu^t}(q_2,q_1) s_{\varphi}(q_2^{-\rho}q_1^{-\nu^t}) = \sum_{\varphi} c_{\lambda,\mu}^{\varphi} Z_{\varphi} \,.
\]
Replacing this in (\ref{eq67}) gives
\begin{eqnarray*}
\bar{\mathcal{P}}^{\lambda\mu}(\text{Hopf})&=&(-1)^{|\lambda|+|\mu|} \left(\frac{q_1}{q_2}\right)^{|\lambda||\mu|} \left(Q^{-1}\sqrt{\frac{q_1}{q_2}}\right)^{\frac{|\lambda|+|\mu|}{2}}\times\frac{Z_{\lambda\mu}}{Z_{\emptyset\emptyset}}=\\
&=&(-1)^{|\lambda|+|\mu|} \left(\frac{q_1}{q_2}\right)^{|\lambda||\mu|} \left(Q^{-1}\sqrt{\frac{q_1}{q_2}}\right)^{\frac{|\lambda|+|\mu|}{2}}\sum_{\varphi} c_{\lambda,\mu}^{\varphi}\frac{Z_{\varphi}}{Z_{\emptyset\emptyset}}=\\
&=& \left(\frac{q_1}{q_2}\right)^{|\lambda||\mu|} \sum_{\varphi} c_{\lambda,\mu}^{\varphi} (-1)^{|\varphi|} \left(Q^{-1}\sqrt{\frac{q_1}{q_2}}\right)^{\frac{|\varphi|}{2}} \frac{Z_{\varphi}}{Z_{\emptyset\emptyset}}=\\
&=&  \left(\frac{q_1}{q_2}\right)^{|\lambda||\mu|} \sum_{\varphi} c_{\lambda,\mu}^{\varphi} \bar{\mathcal{P}}^{\varphi}(\unknot) \,.
\end{eqnarray*}
Equivalently, in the knot theory variables $(\a,q,t)$ we found the following simple formula
for the superpolynomial of the Hopf link expressed in terms of that of the unknot:
\begin{equation}\label{hopfun}
\bar{\mathcal{P}}^{\lambda\mu}(\text{Hopf}) \; = \;
t^{2|\lambda||\mu|} \sum_{\varphi} c_{\lambda,\mu}^{\varphi} \bar{\mathcal{P}}^{\varphi}(\unknot) \,.
\end{equation}

Thus, in order to compute the unreduced superpolynomial of the Hopf link,
it suffices to compute the superpolynomial of the unknot from (\ref{eq67}).

\subsection{Unreduced colored HOMFLYPT polynomial and homology of the unknot}

Below we give the results for the unknot derived from (\ref{eq67}). The notations and computations are summarized in Appendix \ref{compun}.

The quantum $sl(N)$ invariant (that is, $\a=q^N$ specialization of the colored HOMFLY polynomial) is given by:
\begin{eqnarray*}
\bar{{P}}^{\lambda}(\unknot)(\a=q^N,q)&=&q^{-2\sum_{x\in\lambda} c(x)}\left[\!\!\!\begin{array}{c} N\\\lambda^t \end{array}\!\!\! \right]=\\
&=&q^{-2(n(\lambda^t )-n(\lambda ) )} \left[\!\!\!\begin{array}{c} N\\\lambda^t \end{array} \!\!\!\right]= q^{-\kappa(\lambda)} \left[\!\!\!\begin{array}{c} N\\\lambda^t \end{array}\!\!\! \right] \,.
\end{eqnarray*}
In particular, for the $r$-th symmetric representation $S^r$ we find
\be
\label{polsim}
\bar{{P}}^{S^r}(\unknot)(\a=q^N,q) \; = \; q^{-r(r-1)} \left[\!\!\begin{array}{c} N+r-1\\ r \end{array}\!\!\right] \,,
\ee
whereas for the anti-symmetric representation $R = \Lambda^r$ we have
\be
\bar{{P}}^{\Lambda^r}(\unknot)(\a=q^N,q) \; = \; q^{r(r-1)} \left[\!\!\begin{array}{c} N\\ r \end{array}\!\!\right] \,.
\ee

The two-variable polynomial $\bar{{P}}^{\lambda}(\unknot)(\a,q)$ can be obtained from the above expressions
by replacing $q^N$ with $\a$ and the $q$-binomial coefficients by two-variable polynomials in the following way:
\begin{equation}\label{pom1}
 \!\!\!\!\!\! \left[\!\!\begin{array}{c}N \\ r-j \end{array}\!\!\right]  \leftrightarrow  \frac{(-1)^{r-j} \a^{r-j} q^{r-j}}{(1-q^2)(1-q^4)\ldots(1-q^{2r})} \sum_{l=0}^r { (-1)^l \a^{-2l} q^{l(r-j-1)} \left[\!\!\begin{array}{c}r-j \\l\end{array}\!\!\right]  (1-q^{2(r-j+1)})\cdots (1-q^{2r})   } \,,
\end{equation}
where the left hand side is the $\a=q^N$ specialization of the right hand side.
This last formula follows from
\[
\prod_{i=0}^{n-1} {(1+q^{2i}z)} \; = \; \sum_{j=0}^n { q^{j(n-1)}  \left[\!\!\begin{array}{c} n \\ j \end{array}\!\!\right]  z^j   } \,.
\]

In particular, for the symmetric representation we have:
\begin{equation}\label{sim1}
\bar{P}^{S^r}(\unknot)(\a,q)=\frac{(-1)^r\a^rq^r}{(1-q^2)(1-q^4)\ldots(1-q^{2r})} q^{-2r(r-1)}\sum_{l=0}^r { (-1)^l \a^{-2l} q^{l(r-1)} \left[\!\!\begin{array}{c}r \\l\end{array}\!\!\right]  q^{2(r-l)(r-1)}    } .
\end{equation}

Now, the formula for the $\a=q^N$ specialization of the $S^r$-colored superpolynomial
for the unknot is obtained by using the following quantum binomial coefficients formula:
\begin{equation}
\left[\!\!\begin{array}{c} N+r-1 \\ r \end{array}\!\!\right] \; = \; q^{r(r-1)} \sum_{j=0}^{r-1} q^{-j(N+r-1)} \left[\!\!\begin{array}{c} r-1 \\ j \end{array}\!\!\right] \left[\!\!\begin{array}{c} N \\ r-j \end{array}\!\!\right] \,.
\end{equation}
Then, the Poincar\'e polynomial of the $S^r$-colored $sl(N)$ homology of the unknot
is obtained by adding a factor $t^{-2j}$ in every summand in the above expression for the quantum
binomial coefficient in (\ref{polsim}):
\begin{eqnarray}
\bar{\mathcal{P}}^{sl(N),S^r}(\unknot)(q,t) & = &
\bar{\mathcal{P}}^{S^r}(\unknot)(\a=q^N,q,t) = \label{simgl} \\
& = & \sum_{j=0}^{r-1} q^{-j(N+r-1)} \left[\!\!\begin{array}{c} r-1 \\ j \end{array}\!\!\right]
\left[\!\!\begin{array}{c} N \\ r-j \end{array}\!\!\right]t^{-2j} \,. \nonumber
\end{eqnarray}
Note that the corresponding homology $\bar{\mathcal{\CH}}^{sl(N),S^r}(\unknot)$ is finite-dimensional.

We list some particular instances of (\ref{simgl}) for small $r$:
\begin{eqnarray}
\bar{\mathcal{P}}^{sl(N),S^1}(\unknot)&=& [N] \,,\\
\bar{\mathcal{P}}^{sl(N),S^2}(\unknot)&=& \left[\!\!\begin{array}{c} N \\ 2 \end{array}\!\!\right] + q^{-(N+1)}[N]t^{-2} \,,\\
\bar{\mathcal{P}}^{sl(N),S^3}(\unknot)&=&  \left[\!\!\begin{array}{c} N \\ 3 \end{array}\!\!\right] + q^{-(N+2)}[2] \left[\!\!\begin{array}{c} N \\ 2 \end{array}\!\!\right]t^{-2} + q^{-2(N+2)}[N]t^{-4} \,.
\end{eqnarray}
Specifying further the value of $N$, one finds the following expressions:
\begin{eqnarray*}
\bar{\mathcal{P}}^{sl(2),S^2}(\unknot)&=&1+q^{-2}t^{-2}+q^{-4}t^{-2},\\
\bar{\mathcal{P}}^{sl(3),S^2}(\unknot)&=&q^2+1+q^{-2}+q^{-2}t^{-2}+q^{-4}t^{-2}+q^{-6}t^{-2},\\
\bar{\mathcal{P}}^{sl(3),S^3}(\unknot)&=&1+q^{-2}t^{-2}+2q^{-4}t^{-2}+2q^{-6}t^{-2}+q^{-8}t^{-2}+q^{-8}t^{-4}+q^{-10}t^{-4}+q^{-12}t^{-4},\\
\bar{\mathcal{P}}^{sl(4),S^2}(\unknot)&=&q^4+q^2+2+q^{-2}+q^{-4}+q^{-2}t^{-2}+q^{-4}t^{-2}+q^{-6}t^{-2}+q^{-8}t^{-2}.
\end{eqnarray*}
The expression for the whole triply-graded superpolynomial is obtained from (\ref{simgl}) by using (\ref{pom1}).\\

In the case of anti-symmetric representations, the entire homology of the unknot is concentrated
in the homological degree zero, and thus the $\Lambda^r$ superpolynomial of the unknot coincides
with its $\Lambda^r$-colored HOMFLY polynomial:
\begin{equation}
\bar{\mathcal{P}}^{sl(N),\Lambda^r}(\unknot) \; = \; q^{r(r-1)}  \left[\!\!\begin{array}{c} N \\ r \end{array}\!\!\right] \,.
\end{equation}

%%%%%%%%%%%%%%%%%%%%%%%%%%%%%%%%%%%%%%%%%%%%%%%%%%%%%%%%%%%%%%%%%%%%%

\subsection{Comparison with other approaches}
\label{sec:comparison}

Much of the present paper is devoted to exploring the structure --- motivated from physics --- of the colored knot homology,
namely its reduced version. A combinatorial or group theoretic definition of such theory is still waiting to be discovered.
However, in the case of the unreduced theory, which we sketched in this section, there have been several attempts
to define the colored knot homology, especially in low rank.
Therefore, we conclude this section with a brief comparison to other approaches.

Unfortunately, the structure of the colored differentials becomes more obscure (alternatively, more interesting!) in the unreduced version of
the colored knot homology.\footnote{This is familiar from the ordinary, non-colored knot homology \cite{DGR,GWalcher}.}
This, in part, is the reason why we kept our discussion here very brief, relegating a more thorough analysis to future work.
Another reason, which will become clear in a moment, is that even a quick look at the unknot
exposes a number of questions that need to be understood in order to relate and unify different formulations:

\begin{itemize}

\item
singularities in moduli spaces (of BPS configurations):
$\dim \CH_{\text{BPS}} < \infty$ versus $\dim \CH_{\text{BPS}} = \infty$

\item
framing dependence in the colored knot homology

\item
colored homological invariants versus cabling

\item
analog of wall crossing phenomena in mathematical formulations of colored knot homologies

\item
the role of the ``preferred direction'' in the combinatorial formulation based on 3d partitions

\item
proper interpretation of formal expressions, or
\be
\frac{1-q^2}{1-q^2} = (1-q^2)(1 + q^2 + q^4 + q^6 + \ldots) \qquad \text{versus} \qquad \frac{1-q^2}{1-q^2}=1
\label{susyratio}
\ee

\end{itemize}

\noindent
In addition, each formulation typically involves individual choices and subtleties, which may also affect the form of the answer.
In fact, even the total dimension of the colored homology may depend on some (or, perhaps, all)
of these choices.\footnote{We hope that at least some of these delicate aspects are washed away when one passes
to the reduced theory, as it happens in the non-colored case. This is one of the reasons why in the present paper
we mainly consider the reduced homology.}

While good understanding of these aspects is still lacking, many approaches to colored knot homology seem to agree
on one general feature: the unreduced $sl(N)$ homology has finite support only for certain sufficiently small representations.
For example, in \cite[eq. (67)]{GIKV} this corresponds to the fact that for general representations
there is no way to clear the denominators.
This should be compared with the fundamental representation of $sl(N)$,
where every existent approach leads to a homology with finite support.
The simplest example that belongs to the ``grey territory'' is the second symmetric representation $R=S^2$ of $sl(N)$.
For $N=2$, this corresponds to the adjoint representation of $sl(2)$ and, as we saw in \eqref{HHunknot3dim},
physics realizations \cite{GSV,fiveknots} lead to a 3-dimensional knot homology
$\bar{\CH}^{sl(2), \tableau{2}} (\unknot)$ categorifying the colored Jones polynomial of the unknot,
\be
\bar P^{\tableau{2}}_2 (\unknot) \; = \; [3] \; = \; q^{-2} + 1 + q^2 \,.
\label{coloredJunknot}
\ee
On the other hand, some mathematical formulations lead to a theory with infinite support
(which can be attributed to several gaps in the present understanding and the above-listed questions).
For example, fixing\footnote{We thank E.~Gorsky for pointing this out.} a typo in \cite[Proposition 3.4]{Webster},
one finds the following candidate for the Poincar\'e polynomial of the colored unknot homology:
\be
\bar \CP^{\tableau{2}}_2 (\unknot) \; = \; q^{-2} t^2 + 1 + q^2 t^{-2} + \frac{q^{-2} + q^{-2} t}{1-t^2 q^{-4}} \,.
\label{PWebster}
\ee
The structure of the corresponding homology theory is clear: the first three terms reproduce (upon specializing to $t=-1$)
the colored Jones polynomial \eqref{coloredJunknot} and the quotient in the last term corresponds to
the infinite-dimensional contribution to the homology, all of which disappears upon taking the Euler characteristic.

Similar structure emerges in other frameworks,
in particular in approaches based on categorification of the Jones-Wenzl projectors.
The Jones-Wenzl projectors appear in decomposing the finite dimensional representations
of the quantum group $U_q (\frak{sl}_2)$ and, as such, play a key role in the definition
and computation of quantum group invariants of knots and 3-manifolds.
Several ways to categorify the Jones-Wenzl projectors have been proposed in the literature,
{\it e.g.} the topological categorification \cite{CKrushkal}
and the Lie theoretic categorification \cite{Stroppel} which agree (up to Koszul duality).
In particular, the latter approach leads to a theory that categorifies \eqref{coloredJunknot}
by replacing the middle term with infinite-dimensional homology whose Poincar\'e polynomial equals
\be
\bar \CP^{\tableau{2}}_2 (\unknot) \; = \; q^{-2} + \frac{1}{[2] [2]} (q+q^{-1})^2 + q^2 \,,
\label{PStroppel}
\ee
where $[2] = q+q^{-1}$ and the authors of \cite{Stroppel} instruct us to interpret $\frac{1}{[2]}$
as a power series $q - q^3 + q^5 - q^7 + \ldots$. This power series is familiar to physicists as
a trace (``partition function'') over the infinite-dimensional Hilbert space
$\CH_{\text{Bose}} = H^* (\cp^{\infty}) = \C [x]$
of a harmonic oscillator / single boson,
\be
P_{\text{Bose}} \; = \; \frac{1}{1-q^2} = 1 + q^2 + q^4 + q^6 + \ldots \,.
\label{ZBose}
\ee
Partition function of a single fermion has a similar form, except that fermions contribute to
the numerator instead of the denominator. Indeed, the trace over a two-dimensional Hilbert space
of a single fermion looks like
\be
P_{\text{Fermi}} \; = \; 1-q^2 \,,
\label{ZFermi}
\ee
in agreement with a well-known fact that contributions of bosons and fermions cancel each other, {\it cf.} \eqref{susyratio}.
Therefore, instead of canceling the ratio in the middle term of \eqref{PStroppel},
the authors of \cite{Stroppel} instruct us to interpret it as a Hilbert space of two bosons and two fermions.
Note, due to the presence of bosonic states this Hilbert space is infinite-dimensional,
as opposed to a much smaller, finite-dimensional space that one might infer by simplifying the ratio.
Similarly, \eqref{PWebster} contains one boson
(due to the factor $\frac{1}{1-t^2 q^{-4}}$ in the last term), {\it etc.}

If this, however, is the proper interpretation of \eqref{PStroppel}, then one immediately runs into
a general question of how to interpret formal expressions like \eqref{susyratio} and when to clear denominators.
The answer to this question will certainly affect many calculations of Poincar\'e polynomials,
in particular calculations based on \cite[eq. (67)]{GIKV} that has non-trivial numerators and denominators,
as well as similar calculations in other frameworks.

A novel physical framework that appears to be closely related to knot homology
is the so-called ``refined Chern-Simons theory.''
Although Lagrangian definition of this theory is not known at present,
its partition function was conjectured \cite{AShakirov} to compute topological invariants
of knots and 3-manifolds that preserve an extra rotation symmetry.
This includes torus knots and Seifert 3-manifolds.
The rotation symmetry gives rise to an extra quantum number, so that for torus knots
and Seifert 3-manifolds the refined Chern-Simons theory leads to a striking prediction:
the space \eqref{HHBPS} is quadruply-graded rather than triply-graded in these cases.

In simple examples, the fourth grading (coming from the extra rotation symmetry of a 3-manifold)
is determined by the other three gradings \eqref{aqtgradings}.
It would be interesting to study under which conditions this happens; when it does, the partition function
of the $SU(N)$ refined Chern-Simons theory computes the specialization of the superpolynomial to $\a = q^N$.
Assuming this is the case for the unknot colored by the second symmetric representation,
the $SU(2)$ refined Chern-Simons theory gives:
\be
\bar \CP^{\tableau{2}}_2 (\unknot) \; = \;
- \frac{(q^2 t + q^{-2})}{(q - q^{-1})} \frac{(q^4 t^2 + q^{-2} t^{-1})}{(q^3 t^2 - q^{-1})} \,.
\label{PrefinedCS}
\ee
The corresponding Hilbert space contains at least two ``bosons'' (due to two factors in the denominator of \eqref{PrefinedCS})
and, therefore, leads to a version of colored homology with infinite support.

In our quick tour through different ways of categorifying the colored Jones polynomial of the unknot \eqref{coloredJunknot}
we saw theories with finite support as well as theories with infinite support, in fact, of different kind
(with different number of ``bosons'' / factors in the denominator).
One would hope that all these theories correspond to different choices (of framing, chamber, regularization, ... )
and with a proper understanding of the above-mentioned issues could be unified in a single framework.
One piece of evidence that it might be possible
comes from the fact that all physical and geometrical approaches agree when
the corresponding moduli spaces are non-singular, as {\it e.g.} for minuscule representations.
Therefore, we hope to see a much bigger story, only small elements of which have been revealed so far.

%%%%%%%%%%%%%%%%%%%%%%%%%%%%%%%%%%%%%%%%%%%%%%%%%%%%%%

\bigskip
{\it Acknowledgments~}
We would like to thank S.~Cautis, M.~Marino, K.~Schaeffer,
Y.~Soibelman, C.~Stroppel, C.~Vafa, J.~Walcher, and E.~Witten
for valuable discussions.
The work of SG is supported in part by DOE Grant DE-FG03-92-ER40701
and in part by NSF Grant PHY-0757647.
Opinions and conclusions expressed here are those of the authors
and do not necessarily reflect the views of funding agencies.
MS was partially supported by the Funda\c c\~ao para a Ci\^encia e Tecnologia (ISR/IST plurianual funding)
through the programme Programa Operacional Ci\^encia, Tecnologia, Inova\c c\~ao (POCTI) and the POS Conhecimento programme,
co-financed by the European Community fund FEDER.
MS was also partially supported by the Ministry of Science of Serbia, project no. 174012.

%%%%%%%%%%%%%%%%%%%%%%%%%%%%%%%%%%%%%%%%%%%%%%%%%%%%%%
%
%  Appendix

\appendix
\section{Notations}
\label{sec:notation}

$K$ denotes a knot. ${\,\raisebox{-.08cm}{\includegraphics[width=.4cm]{unknot}}\,}$ denotes the unknot.

For every positive integer $r$ we have:

$\bullet$ $\bar{P}^{S^r}_N(K)(q)$ or $\bar{P}^{\overbrace{\tableau2\cdots\tableau1}^r}_N(K)(q)$ denotes the unreduced
one-variable polynomial quantum invariant of $K$, labelled with the $r$-th symmetric representation of $sl(N)$.

$\bullet$ $P^{S^r}_N(K)(q)$ or $P^{\overbrace{\tableau2\cdots\tableau1}^r}_N(K)(q)$ denotes the reduced (a.k.a. normalized)
one-variable polynomial quantum invariant of $K$, labeled by the $r$-th symmetric representation of $sl(N)$.
It is obtained from the unnormalized polynomial $\bar{P}^{S^r}_N(K)(q)$ by
\[
P^{S^r}_N(K)=\frac{\bar{P}^{S^r}_N(K)}{\bar{P}^{S^r}_N({\,\raisebox{-.08cm}{\includegraphics[width=.4cm]{unknot}}\,})} \,,
\]
so that $P^{S^r}_N({\,\raisebox{-.08cm}{\includegraphics[width=.4cm]{unknot}}\,})=1$.\\

$\bullet$ $P^{S^r}(K)(a,q)$ or $P^{\overbrace{\tableau2\cdots\tableau1}^r}(K)(\a,q)$ denotes the reduced two-variable colored HOMFLY polynomial of $K$. The normalization is  $P^{S^r}({\,\raisebox{-.08cm}{\includegraphics[width=.4cm]{unknot}}\,})=1$. In particular
\[
P^{S^r}(K)(\a=q^N,q)=P^{S^r}_N(K)(q)
\]

$\bullet$ $\CH^{S^r}(K)$ or $\CH^{\overbrace{\tableau2\cdots\tableau1}^r}(K)$ denotes a reduced triply-graded homology theory
$\CH^{S^r}_{i,j,k}(K)$ that categorifies the two-variable colored HOMFLY polynomial $P^{S^r}(K)$:
\[
\chi(\CH^{S^r}(K))=P^{S^r}(K).
\]

The homological grading of $\CH^{S^r}_{i,j,k}(K)$ is its third grading, and its (doubly-graded) Euler characteristic is given by:
\[
\chi(\CH^{S^r}(K))=\sum_{i,j,k} (-1)^k q^it^j (\dim \CH^{S^r}_{i,j,k}(K)).
\]

$\bullet$ $\CP^{S^r}(K)$ or $\CP^{\overbrace{\tableau2\cdots\tableau1}^r}(K)$ denotes the Poincar\'e polynomial of
$\CH^{S^r}(K)$, {\it i.e.}
\[
\CP^{S^r}(K)(\a,q,t) \; = \; \sum_{i,j,k} \a^iq^jt^k (\dim \CH^{S^r}_{i,j,k}(K)).
\]
Specialization to $t=-1$ gives $P^{S^r}(K)$:
\[
\CP^{S^r}(K)(\a,q,t=-1) \; = \; P^{S^r}(K)(\a,q) \,.
\]
We also call $\CP^{S^r}(K)$ the $S^r$-\textit{colored superpolynomial}.\\

$\bullet$ $\CH^{sl(N),S^r}(K)$ denotes the reduced doubly-graded homology theory categorifying $P^{S^r}_N(K)$.\\

$\bullet$ $\CP^{S^r}_N(K)$ denotes the Poincar\'e polynomial of $\CH^{sl(N),S^r}(K)$. In particular
\[
\CP^{S^r}_N(K)(q,t) \; = \; \sum_{j,k} q^jt^k (\dim \CH^{sl(N),S^r}_{j,k}(K)) \,.
\]
\[
\CP^{S^r}_N(K)(q,t=-1) \; = \; P^{S^r}_N(K)(q) \,.
\]

\vskip 0.2cm

$\bullet$ The corresponding polynomials and homologies for the anti-symmetric representations are denoted in the same way with $S^r$ replaced by
$\Lambda^r$.

\vskip 0.2cm

$\bullet$ Unreduced versions of the homology and polynomials are denoted by putting a bar.

\vskip 0.2cm

All (tri-)degrees are $(\a,q,t)$-degrees.

\begin{remark} \label{rem1}
\textbf{Conventions for the superpolynomial in the vector representation: } The case $r=1$ corresponds to the (ordinary) HOMFLY polynomial and the Khovanov-Rozansky homology. The corresponding superpolynomial, together with the structure of the triply-graded homology was studied in \cite{DGR}.

Here we shall use slightly different conventions: in the superpolynomial, we replace $\a$ and $q$ from \cite{DGR} with $\a^{1/2}$ and $q^{1/2}$, respectively. For example, the superpolynomial of the trefoil knot becomes:
\begin{equation}\label{p31}
\CP^{\tableau1} (3_1)=\a q^{-1}+\a qt^2+\a^2t^3.
\end{equation}

Also, by $S(K)$, or just $S$, we mean half of the value of $S(K)$ from \cite{DGR}.

This way, the degrees of canceling differentials $d_1$ and $d_{-1}$ from \cite{DGR} become $(-1,1,-1)$ and $(-1,-1,-3)$, respectively.
Also, from now on we denote these two differentials by $d_1^{\tableau1}$ and $d^{\tableau1}_{-1}$, respectively.
\end{remark}

%%%%%%%%%%%%%%%%%%%%%%%%%%%%%%%%%%%%%%%%%%%%%%%%%%%%%%%%%%%%%%%%%%%%%%%%%%%

\section{Kauffman and $S^2$ homologies of the knots $8_{19}$ and $9_{42}$}
\label{sec:942}

In this appendix we compute the $S^2$ and Kauffman homologies of $8_{19}$ and $9_{42}$.
Since both knots are homologically thick knots, these computations give highly non-trivial examples
of our main Conjecture \ref{con2} (as well as Conjecture 2 of \cite{GWalcher} for the Kauffman homology),
especially because the size of both homologies is rather large and because they have to obey
a large list of structural properties described in sections \ref{sec:reduced} and \ref{sec:mirror}.

Before we start our computations, we point out that knot $9_{42}$ here is the mirror image of $9_{42}$ from \cite{KAt}.
The superpolynomial of $9_{42}$ is given by\footnote{Note that the $\a$ and $q$ gradings
that we are using in this paper are half of those from \cite{DGR}, see Remark \ref{rem1}} \cite{DGR}:
\be\label{942prva}
\CP^{\tableau1}(9_{42})(\a,q,t)=\a (q^{-1}t^2+qt^4)+(q^{-2}t^{-1}+1+2t+q^2t^3)+\a^{-1}(q^{-1}t^{-2}+q).
\ee
In particular, the reduced $sl(2)$ Khovanov homology of $9_{42}$ is:
\begin{eqnarray*}
Kh(9_{42}) (q,t) & = & \CP^{\tableau1}(9_{42})(\a=q^2,q,t) = \\
& = & q^{-6}t^{-2} +q^{-4}t^{-1} + q^{-2}+1 + 2t +q^2t^2 + q^4t^3 +q^6t^4 \,.
\end{eqnarray*}
The $S$ invariant is $S(9_{42})=0$.
Moreover, the $\delta$-grading of a generator $x$ of the homology $\CH^{\tableau1}(K)$ in our conventions is given by:
\[
\delta(x)=t(x)-2 \a (x)-q(x).
\]
All generators of the homology of thin knots have the same value of the $\delta$-grading.
However, for $9_{42}$, the generator $1(=\a^0q^0t^0)$ has $\delta$-grading 0,
while the remaining 8 generators have $\delta$-grading equal to $-1$.\\

As for $8_{19}$, here it is the mirror image of $8_{19}$ from the Knot Atlas \cite{KAt}.
This knot is also known as the positive (3,4)-torus knot.
Its superpolynomial is given by:
\begin{eqnarray*}
\CP^{\tableau1}(8_{19})(\a,q,t)&=&\a^3q^{-3}+\a^4q^{-2}t^3+\a^3q^{-1}t^2+\a^4t^5+\a^3qt^4+\a^4q^{2}t^7+\a^4q^{3}t^6+ \\
&&+\a^5t^8+\a^4q^{-1}t^5+\a^4q^{1}t^7+\a^3t^4.
\end{eqnarray*}
The first seven generators have $\delta$-degree equal to $-3$, while the remaining four have $\delta$-grading equal to $-2$.\\

Before explaining the result for the $S^2$-homology, we first consider the Kauffman homologies.

\subsubsection{Kauffman homology of $8_{19}$ and $9_{42}$}\label{Kauf}

The Kauffman polynomial of $9_{42}$ can be written as:
\be
F(9_{42})(\a,q)=1+ (1-\a^{-1}q)(1+\a^{-1}q^{-1})(1-\a^{-2}) \cdot \a^2(q^{-6} - q^{-4} +q^{-2} +q^2 - q^4 +q^6).
\ee

The Kauffman homology of $9_{42}$ that we have computed has 209 generators.
We present its Poincar\'e polynomial in a structured form:
\begin{eqnarray}
\mathcal{P}^{\mathrm{Kauff}}(9_{42})(\a,q,t)&=&1+ (1+\a^{-1}qt^{-1})(1+\a^{-1}q^{-1}t^{-2})(1+\a^{-2}t^{-3}) \times \nonumber\\
&&\times \{\a^2(q^{-6}t^2+q^{-4}t^3+q^{-2}t^4+q^2t^6+q^4t^7+q^6t^8)+ \nonumber \\
&&\quad\quad+(1+t)\left[(\a^3+\a t^{-3})(q^{-3}t^4+q^{-1}t^5+qt^6+q^3t^7)+2\a^2t^4\right] \}  \nonumber
\end{eqnarray}

The Kauffman polynomial of $8_{19}$ can be written as:
\[
F(8_{19})(\a,q)=(\a^6-\a^8)(q^{-6}+q^{-2}+1+q^2+q^6)+(\a^7-\a^9)(q^{-5}-q^5)+\a^{10}.
\]

The Kauffman homology of $8_{19}$ that we have computed has 89 generators.
Its Poincar\'e polynomial is given by:
\begin{eqnarray*}
\mathcal{P}^{\mathrm{Kauff}}(8_{19})(\a,q,t)&=&
(\a^6+\a^8t^3)(q^{-6}+q^{-2}t^2+t^4+q^2t^4+q^6t^6)+\\
&&+(\a^7+\a^9t^3)(q^{-5}t^2+q^5t^7)+\a^{10}t^{10}+\\
&&+(1+t^{-1})(\a^7q^{-1}t^5+\a^7q^{1}t^6+\a^8t^7+\a^9q^{-1}t^8+\a^9q^{1}t^9)+\\
&&+(1+t^{-1})(1+\a^{-1}qt^{-1})(1+\a^{-2}t^{-3})(1+\a^{-1}q^{-1}t^{-2})\times\\
&&\quad\quad \times (\a^{13}q^{-1}t^{14}+\a^{13}q^1t^{15}+\a^{11}q^{-3}t^{10}+\a^{11}q^3t^{13}).
\end{eqnarray*}

\begin{figure}[htb]
\centering
\includegraphics[width=5.0in]{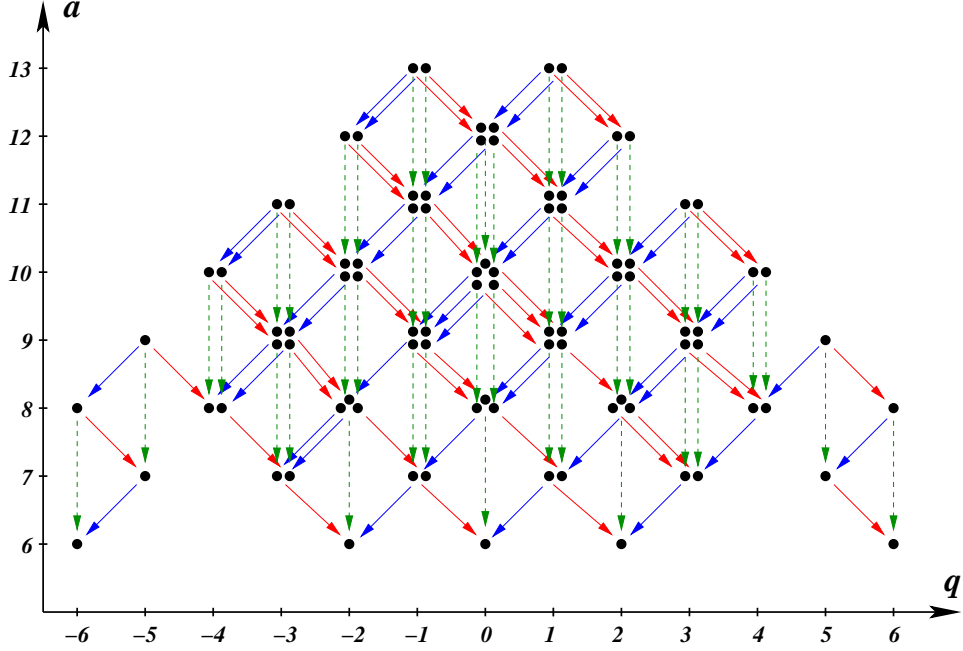}
\caption{The reduced Kauffman homology of the knot $T_{3,4} = 8_{19}$.
To avoid clutter, we show only canceling differentials $d_2$, $d_1$ and $d_0$
(represented by red, green, and blue arrows, respectively).}
\label{fig:kauffman819}
\end{figure}

Both results meet the desired properties of the Kauffman homology (see Section 6 of \cite{GWalcher}):

\begin{itemize}

\item Specialization to $t=-1$ gives the Kauffman polynomial:
 \[
\mathcal{P}^{\mathrm{Kauff}}(\a,q,t=-1)=F(\a,q).
 \]

\item
There exist three canceling differentials $d_2$, $d_1$ and $d_0$ of degrees $(-1,1,-1)$, $(-2,0,-3)$ and $(-1,-1,-2)$, respectively.
Indeed, from the form we write $\CP^{\mathrm{Kauff}}(9_{42})$, it is obvious that only $1=\a^0q^0t^0$
survives in the homology with respect to any of these differentials.
As for $\CH^{\mathrm{Kauff}}(8_{19})$, the surviving generators for $d_2$, $d_1$ and $d_0$ respectively
have degrees $(6,-6,0)$, $(12,0,12)$ and $(6,6,6)$, as expected since the $S$-invariant for $8_{19}$ is $S(8_{19})=6$.

\item
There exist two universal differentials $d_{\rightarrow}$ and $d_{\leftarrow}$ of degrees $(0,2,1)$ and $(0,-2,-1)$,
such that the homology with respect to these differentials is isomorphic (up to regrading) to the triply-graded HOMFLY homology. More precisely, they satisfy:
\[
(\CH^{\mathrm{Kauff}}, d_{\rightarrow}) =  \CP^{\tableau1} (\a^2q^{-2},q^2,t),
\]
\[
(\CH^{\mathrm{Kauff}}, d_{\leftarrow})  = \CP^{\tableau1}(\a^2q^2t^2,q^2,t).
\]
Again, it is straightforward to check that such differentials exist in both homologies $\CH^{\mathrm{Kauff}}(9_{42})$ and $\CH^{\mathrm{Kauff}}(8_{19})$.

%\item
%There exists a differential $d_4$ of degree $(-1,3,-1)$, such that the $\a=q^3$ specialization of the homology %$(\CH^{\mathrm{Kauff}}(9_42), d_4)$ is isomorphic to  the homology $\CH^{so(4),V}(9_{42})$. The Poincar\'e polynomial of the latter %homology can be computed via Khovanov ($sl(2)$) homology in a following way:
%\[
%HSO_4(9_{42})(q,t)= (Kh(9_{42})(q,t))^2.
%\]
%In particular, it has 81 terms, and it is straightforward to check that there exists a differential $d_4$ in %$\CH^{\mathrm{Kauff}}(9_42)$ that has the wanted properties.

\item
There exists a  differential $d_{-2}$ of degree $(-1,-3,-3)$, such that the $\a=q^{-3}$ specialization of the homology $(\CH^{\mathrm{Kauff}}, d_{-2})$ is isomorphic to  the homology $\CH^{sp(2),V}$.

Moreover, the triply-graded version holds: the Poincar\'e polynomial of the triply-graded homology $(\CH^{\mathrm{Kauff}}, d_{-2})$ is equal to
$t^S R(a^{1/2}q^{-1/2},t)$, where $R(q,t)=\CP^{\tableau1}(\a=q^2,q,t)$. This is true for $8_{19}$, $9_{42}$ and for all prime knots with up to 6 crossings.

Note that this generalizes and corrects the value predicted in \cite{GWalcher}. We also note that $(\CH^{\mathrm{Kauff}}, d_{-2})$ is significantly smaller than $\CH^{\mathrm{Kauff}}$: for $8_{19}$ it has only 11 generators, and for $9_{42}$ it has only 9 generators.\\

\item
Finally, although not explicitly stated in \cite{GWalcher}, the Kauffman homology enjoys the symmetry $q \leftrightarrow q^{-1}$:
\[
\CH^{\mathrm{Kauff}}_{i,j,*} \; \cong \; \CH^{\mathrm{Kauff}}_{i,-j,*} \,.
\]

\end{itemize}

\subsubsection{$S^2$-colored homologies of $8_{19}$ and $9_{42}$}

In order to present the Poincar\'e polynomial of the $S^2$-colored homology of $9_{42}$ in a nice form
and to show that all the expected properties are satisfied, we write it in the following structured form:
\begin{eqnarray*}
&& \CP^{\tableau2}(9_{42})(\a,q,t) = \\
&& = \{1 + (1+\a^{-1}qt^{-1})(1+\a^{-1}q^{-2}t^{-3})(1+\a^{-1}q^{-3}t^{-3})(1+\a^{-1}t^{-1})(\a^2q^6t^8+\a^2t^4)\}+\\
&&\quad +(1+q)(1+\a^{-1}qt^{-1})(1+\a^{-1}q^{-2}t^{-3})(1+\a^{-1}q^{-3}t^{-3})(\a^2 q^2t^6+\a q^3t^5)+\\
&&\quad +(1+q)(1+\a^{-1}qt^{-1})(1+\a^{-1}q^{-2}t^{-3})(\a q^{-1}t^2+\a qt^4)+\\
&&\quad +(1+q)(1+\a^{-1}qt^{-1})(1+\a^{-1}q^{-2}t^{-3})(1+\a^{-1}q^{-3}t^{-3})\times\\
&&\quad \quad\quad\quad\times(\a^2q^2t^6+ \a q^3t^6 +\a q^3t^5 + 2\a qt^5+ \a t^4 +qt^4 +t^3) +\\
&&\quad +(1+q)(1+\a^{-1}qt^{-1})(1+\a^{-1}q^{-2}t^{-3})(1+\a^{-1}q^{-3}t^{-3})(1+t^{-1})\times\\
&&\quad \quad\quad\quad\times(\a^2q^4t^8+\a q^4t^7+\a^2qt^6+\a q^2t^5+q^3t^5+\a t^4+t^3).
\end{eqnarray*}

Similarly, for $8_{19}$ we have
\begin{eqnarray*}
&& \CP^{\tableau2}(8_{19})(\a,q,t) = \\
&& = \{(1+\a^{-1}qt^{-1})(1+\a^{-1}q^{-2}t^{-3})(1+\a^{-1}q^{-3}t^{-3})(\a^{10}q^8t^{16}+\a^9q^8t^{15})+\\
&&\quad  +(1+\a^{-1}qt^{-1})(1+\a^{-1}q^{-2}t^{-3})(\a^8q^{-1}t^6+\a^8q^5t^{10}+\a^8q^{11}t^{14})+\a^6q^{-6} +\\
&&\quad  +(1+\a^{-1}qt^{-1})(\a^7q^{-4}t^{3}+\a^7q^{-1}t^{5}+\a^7q^{2}t^{7}+\a^7q^{5}t^{9}+\a^7q^{8}t^{11}+\a^7q^{11}t^{13})\}+\\
&&\quad +(1+q)(1+\a^{-1}qt^{-1})(1+\a^{-1}q^{-2}t^{-3})(\a^8q^2t^8+\a^8q^6t^{10}+\a^8q^8t^{12}+\a^8q^9t^{14})+\\
&&\quad +(1+q)(1+\a^{-1}qt^{-1})(1+\a^{-1}q^{-2}t^{-3})(1+\a^{-1}q^{-3}t^{-3})(\a^9q^3t^{11}+\a^9q^6t^{13}+\a^9q^9t^{15}).
\end{eqnarray*}

The two homologies from above and their mirror images satisfy a large part
of the properties of the $S^2$-colored homology from Conjecture \ref{con2} and of the $\Lambda^2$ homology from Section~\ref{sec:mirror}:

\begin{itemize}

\item
There exist canceling differentials $d^{\tableau2}_1$ and $d^{\tableau2}_{-2}$ of degrees $(-1,1,-1)$ and $(-1,-2,-3)$, respectively.
The remaining generator for both differentials is $\a^0 q^0 t^0 = 1$ in the case of $9_{42}$,
whereas for $8_{19}$ the remaining generators have degrees $(6,-6,0)$ and $(6,12,12)$, respectively.

\item
There exists colored differential $d_{2\to 1}$ of degree $(0,1,0)$, such that the homology with respect
to it is equal to $\CP^{\tableau1} (\a^2,q^2,t^2q)$.

\item
$\CP^{\tableau2}(\a,q,t=-1)$ is equal to the $S^2$-colored HOMFLY polynomial.
%We don't have an access to this polynomial of $9_{42}$, but rather only its $\a=q^2$ and $\a=q^3$ specialization from Knot Atlas %\cite{KAt}.
%After direct computation, in terms of the polynomial $R$ from (\ref{ps2}), this condition means
%\begin{eqnarray*}
%R(\a=q^2,q,t=-1)& = &-q^4 + q -1,\\
%R(\a=q^3,q,t=-1)&= &-q^4 + q^2 -1.
%\end{eqnarray*}
%Obviously, our value of $R$ from (\ref{polR}) satisfy these two conditions.

\item
There exists a differential $d^{\tableau2}_{-3}$ of degree $(-1,-3,-3)$,
such that the homology with respect to it is very small. In the case of $9_{42}$ it has only 9 generators:
\begin{equation}
(\CH^{\tableau2}(9_{42}),d^{\tableau2}_{-3}) =
\a (q^{-1}t^2+qt^4)+(q^{-2}t^{-1}+1+2t+q^2t^3)+\a^{-1}(q^{-1}t^{-2}+q),
\label{dmi3}
\end{equation}
while in the case of $8_{19}$ we have
\begin{eqnarray}
(\CH^{\tableau2}(8_{19}),d^{\tableau2}_{-3}) &=& (\a^{6}q^{12}t^{12}+
\a^{7}q^{11}t^{13}+\a^{6}q^{10}t^{10}+\a^{7}q^{9}t^{11}+\a^{6}q^{8}t^{8}+\a^{7}q^{7}t^{9}+ \nonumber\\
&&+\a^{6}q^{6}t^{6}) +(\a^8q^{10}t^{14}+\a^7q^9t^{11}+\a^7q^{11}t^{13}+\a^6 q^{10}t^{10}).\label{dmi3819}
\end{eqnarray}

Note that from the formulas for $\CP^{\tableau2}$, for both knots, the last two lines have a factor $(1+\a^{-1}q^{-3}t^{-3})$ and so the corresponding homology gets canceled automatically by $d^{\tableau2}_{-3}$.
Thus, it is enough to check the above formulas only for the remaining part, which is a straightforward computation.\\

\item
There exists a differential $d^{\tableau2}_0$ of degree $(-1,0,-1)$ such that the homology with respect to it is equal to $\a^S \CP^{\tableau1} (\a,q^2,t)$.\\

\item
There exists a differential $d^{\tableau2}_2$ of degree $(-1,2,-1)$ such that the homology with respect to it, after specializing $\a=q^2$, is isomorphic to
$\CH^{sl(2),\tableau2}$. The latter one is isomorphic to $\CH^{so(3),V}$, where $V$ denotes the vector representation of $so(3)$, and
to obtain its Poincare polynomial we use the result for the Kauffman homology we computed in section \ref{Kauf}. In particular, for both knots we have that
\[
\CP^{\tableau2}_2(q,t)=\CP^{so(3),V}(q,t)=\CP^{\mathrm{Kauff}}(\a=q^2,q,t).
\]

\end{itemize}

Now, the ``mirror image" of $\CH^{\tableau2}$ is also behaving quite well.
To that end, let $\CH^{\tableau{1 1}}$  be a homology obtained from $H^{\tableau2}$ as in Section \ref{sec:mirror}:
\[
\CH^{\tableau{1 1}}_{i,j,k} \cong \CH^{\tableau2}_{i,-j,k'}.
\]
The transformation $k \mapsto k'$ depends also on $\delta'$-grading.\\

Since for $\CH^{\tableau2}$ of $9_{42}$ and $8_{19}$ the properties of the $S^2$-colored homology
listed above are satisfied, it can be easily seen (by ``mirroring" the differentials) that
the mirror homology $\CH^{\tableau{1 1}}$ obtained in this way satisfies the properties
of the anti-symmetric $\Lambda^2$-colored homology:

\begin{itemize}
\item
There exist canceling differentials $d^{\tableau{1 1}}_{-1}$ and $d^{\tableau{1 1}}_{2}$ of degrees
$(-1,-1,-3)$ and $(-1,2,-1)$, respectively.

\item
There exists colored differential $d^{\tableau{1 1}}_{2\to 1}$ of degree $(0,1,0)$, such that the homology with respect to it has
Poincar\'e polynomial equal to $\CP^{\tableau1} (\a^2,q^{4},t^2q^{-1})$.

\item
$\CP^{\tableau{1 1}}(\a,q,t=-1)$ is equal to the $\Lambda^2$-colored HOMFLY polynomial.

\item
There exists a differential $d^{\tableau{1 1}}_3$ of degree $(-1,3,-1)$ such that the homology of $\CH^{\tableau{1 1}}$ with respect to it is isomorphic to $\CH^{\tableau1}$, both specialized to $\a=q^3$.

\noindent

The last property in fact holds even on the level of triply-graded homologies (without specialization $a=q^3$), as can be seen from (\ref{dmi3}) and (\ref{dmi3819}).
We also note that the isomorphism of $(\CH^{\tableau{1 1}},d^{\tableau{1 1}}_3)$ and $\CH^{\tableau1}$ as triply-graded theories, also holds for all prime knots with up to 6 crossings.
\end{itemize}

%%%%%%%%%%%%%%%%%%%%%

\section{$\CH^3$ homology of the figure-eight knot}\label{apb41}
The Poincare polynomial of the $\CH^3$ homology of the figure-eight knot $4_1$ is given by:

\begin{eqnarray*}
\CP^3(4_1)=\CP^{\tableau3}(4_1)&=&1+(1+\a^{-1}qt^{-1})(1+\a^{-1}t^{-1})(1+\a^{-1}q^{-1}t^{-1})\times\\
&&\quad\quad \times (1+\a^{-1}q^{-3}t^{-3})(1+\a^{-1}q^{-4}t^{-3})(1+\a^{-1}q^{-5}t^{-3})\a^3q^6t^6+\\
&&+(1+q+q^2)(1+\a^{-1}qt^{-1})(1+\a^{-1}q^{-3}t^{-3})\a t^2+\\
&&+(1+q+q^2)(1+\a^{-1}qt^{-1})(1+\a^{-1}t^{-1})(1+\a^{-1}q^{-3}t^{-3})(1+\a^{-1}q^{-4}t^{-3})\a^2q^2t^4.
\end{eqnarray*}

This homology categorifies the $\tableau3$-colored HOMFLY polynomial of $4_1$;
\[
\CP^3(4_1)(\a,q,t=-1)=P^{\tableau3}(4_1)(\a,q).
\]

Furthermore, this homology has all of the wanted properties - namely, there exist following differential ons $\CH^3(4_1)$:

\begin{itemize}
\item
canceling differential $d^3_1$ of degree $(-1,1,-1)$, leaving $1=\a^0q^0t^0$ as remaining generator.

\item
canceling differential $d^3_{-3}$ of degree $(-1,-3,-3)$, also leaving $1=\a^0q^0t^0$ as remaining generator.

\item
differential $d^3_{-4}$ of degree $(-1,-4,-3)$, such that
\[
(\CH^3(4_1),d^3_{-4})\cong \CH^1(4_1).
\]

\item
differential $d^3_{-5}$ of degree $(-1,-5,-3)$, such that
\[
(\CH^3(4_1),d^3_{-5})\cong \CH^2(4_1).
\]

\item
vertical colored differential $d^3_{0}$ of degree $(-1,0,-1)$, such that
\[
(\CH^3(4_1),d^3_{0})\cong \CH^1(4_1).
\]
In particular
\[
(\CH^3(4_1),d^3_{0})(\a,q,t)= \CP^1(4_1) (\a,q^3,t).
\]

\item
vertical colored differential $d^3_{-1}$ of degree $(-1,-1,-1)$, such that
\[
(\CH^3(4_1),d^3_{-1})\cong \CH^2(4_1).
\]
In particular
\[
(\CH^3(4_1),d^3_{-1})(\a,q=1,t)= \CP^2(4_1) (\a,q=1,t).
\]

\end{itemize}

%%%%%%%%%%%%%%%%%%%%%%%%%%%%%%%%%%%%%%%%%%%%%%%%%%%%%%%%%

\section{Computation of the unreduced homology of the unknot}
\label{compun}

For a nonnegative integer $N$ we define the quantum dimension $[N]$ to be
\[
[N] \; = \; \frac{q^N-q^{-N}}{q-q^{-1}} \,.
\]

Also
$$
[N]!=[N][N-1]\ldots[1] \,,
$$
and
$$
\left[\!\!\!\begin{array}{c} N\\ k \end{array}\!\!\! \right]=\frac{[N][N-1]\ldots[N-k+1]}{[k]!} \,.
$$
\vskip 0.3cm

For a partition $\lambda=(\lambda_1,\lambda_2,\ldots)$, we set:
\begin{eqnarray*}
|\lambda|&=&\sum_i \lambda_i,\\
n(\lambda)&=&\sum_i {(i-1)\lambda_i},\\
m(\lambda)&=&\sum_i {i\lambda_i},\\
||\lambda||^2&=&\sum_i \lambda_i^2.
\end{eqnarray*}

By $\lambda^t$, we denote the dual (conjugate) partition of $\lambda$. We have
\begin{eqnarray*}
m(\lambda)&=&n(\lambda)+|\lambda|,\\
2m(\lambda)&=& ||\lambda^t||^2+|\lambda|.
\end{eqnarray*}

We also define
\[
\kappa (\lambda):=2(||\lambda||^2 - ||\lambda^t||^2) = 2(n(\lambda^t)-n(\lambda)).
\]

\vskip 0.5cm

We identify a partition and its corresponding Young diagram.

For a box $x=(i,j)\in \lambda$, we define its content and the hook length by:
\begin{eqnarray*}
c(x)&=&j-i,\\
h(x)&=&\lambda_i+\lambda^t_j-i-j+1.
\end{eqnarray*}

For a partition $\lambda$ and positive integer $N$, we set:
\[
\left[\!\!\!\begin{array}{c} N\\\lambda \end{array}\!\!\!\right]:=\prod_{x\in\lambda} \frac{[N-c(x)]}{[h(x)]} \,.
\]
Then we have
\begin{equation}
s_{\lambda}(q^{1-N},q^{3-N},\ldots,q^{N-3},q^{N-1}) \; = \; \left[\!\!\!\begin{array}{c} N\\\lambda^t \end{array} \!\!\!\right].
\end{equation}
Of course, the following holds
\[
s_{\lambda}(1,q^2,q^4,\ldots,q^{2(N-2)},q^{2(N-1)}) \; = \; q^{(N-1)|\lambda|}s_{\lambda}(q^{1-N},q^{3-N},\ldots,q^{N-3},q^{N-1}).
\]

\subsection{Colored HOMFLY polynomial of the unknot}

We compute the specialization of the superpolynomial of the unknot at $q_1=q_2$ ({\it i.e.} at $t=-1$) from
the equation (\ref{eq67}). According to \eqref{hopfun}, this gives the value of the polynomial of the Hopf link as well.
We denote this specialization by $\bar{P}^{\lambda}(\unknot)$, {\it i.e.}
\[
\bar{P}^{\lambda}(\unknot) \; := \; \bar{\mathcal{P}}^{\lambda}(\unknot){|_{q_1=q_2}} \; = \; \bar{\mathcal{P}}^{\lambda}(\unknot){|_{t=-1}} \,.
\]

We also denote $Z'_{\lambda}:=Z_{\lambda}{|_{q_1=q_2}}$, and so
\[
\bar{P}^{\lambda}(\unknot) \; = \; (-1)^{|\lambda|} (Q^{-1})^{\frac{|\lambda|}{2}} \times \frac{Z'_{\lambda}}{Z'_{\emptyset}} \,,
\]

with
\[
Z'_{\lambda}=\sum_{\nu} (-Q)^{|\nu|} s_{\nu} (q_2^{-\rho}) s_{\nu^t} (q_2^{-\rho}) s_{\lambda} (q_2^{-\rho-\nu^t}).
\]

By using the following identity for the Schur functions
\begin{equation}\label{lemica}
s_{\lambda}(q_2^{-\rho-\nu^t}) s_{\nu^t}(q_2^{-\rho}) = q_2^{-\frac{\kappa(\lambda)}{2}} \sum_{\eta} s_{^{\nu^t}\!\!/_{\eta}}(q_2^{-\rho}) s_{^{\lambda^t}\!\!/_{\eta}}(q_2^{-\rho})
\end{equation}

in the expression for $Z'_{\lambda}$, we get
\begin{eqnarray*}
Z'_{\lambda}&=& \sum_{\nu} (-Q)^{|\nu|} s_{\nu} (q_2^{-\rho})  q_2^{-\frac{\kappa(\lambda)}{2}} \sum_{\eta} s_{^{\nu^t}\!\!/_{\eta}}(q_2^{-\rho}) s_{^{\lambda^t}\!\!/_{\eta}}(q_2^{-\rho})=\\
&=& q_2^{-\frac{\kappa(\lambda)}{2}} \sum_{\eta} s_{^{\lambda^t}\!\!/_{\eta}}(q_2^{-\rho})  \sum_{\nu} (-Q)^{|\nu|} s_{\nu} (q_2^{-\rho}) s_{^{\nu^t}\!\!/_{\eta}}(q_2^{-\rho})=\\
&=& q_2^{-\frac{\kappa(\lambda)}{2}} \sum_{\eta} s_{^{\lambda^t}\!\!/_{\eta}}(q_2^{-\rho})  \sum_{\nu} (-Q)^{|\nu|} s_{\nu} (q_2^{-\rho}) \sum_{\varphi} c_{\eta,\varphi}^{\nu^t} s_{\varphi} (q_2^{-\rho})=\\
&=& q_2^{-\frac{\kappa(\lambda)}{2}} \sum_{\eta} s_{^{\lambda^t}\!\!/_{\eta}}(q_2^{-\rho}) \sum_{\varphi}  s_{\varphi} (q_2^{-\rho}) \sum_{\nu} (-Q)^{|\eta|+|\varphi|} c_{\eta,\varphi}^{\nu^t} s_{\nu} (q_2^{-\rho})=\\
&=& q_2^{-\frac{\kappa(\lambda)}{2}} \sum_{\eta} s_{^{\lambda^t}\!\!/_{\eta}}(q_2^{-\rho}) (-Q)^{|\eta|} \sum_{\varphi} (-Q)^{|\varphi|} s_{\varphi} (q_2^{-\rho})  s_{\varphi^t} (q_2^{-\rho}) s_{\eta^t} (q_2^{-\rho})=\\
&=& q_2^{-\frac{\kappa(\lambda)}{2}} \sum_{\eta} (-Q)^{|\eta|} s_{^{\lambda^t}\!\!/_{\eta}}(q_2^{-\rho})  s_{\eta^t} (q_2^{-\rho}) \times Z'_{\emptyset}.
\end{eqnarray*}

Hence, the value of the colored HOMFLY polynomial invariant of the unknot labeled by $\lambda$ is given by:
\begin{equation}\label{unkon}
\bar{P}^{\lambda}(\unknot) \; = \; q_2^{-\frac{\kappa(\lambda)}{2}} \cdot (-1)^{|\lambda|} (Q^{-1})^{\frac{|\lambda|}{2}} \sum_{\eta} (-Q)^{|\eta|} s_{^{\lambda^t}\!\!/_{\eta}}(q_2^{-\rho})  s_{\eta^t} (q_2^{-\rho}) \,.
\end{equation}
In particular, the last formula tells us that (up to an overall factor) $P_{\lambda}$ is a polynomial in $Q$ of degree $|\lambda|$.
Also, at $Q=1$ one finds that $\bar{P}_{\lambda}|_{Q=1}=0$ for $\lambda\ne 0$.\\

\noindent
Written in terms of the knot-theoretical variables $(\a,q)$ the polynomial becomes:
\begin{equation}
\label{unkon2}
\bar{P}^{\lambda}(\unknot)(\a,q) \; = \; q^{-\kappa(\lambda)} \cdot (-1)^{|\lambda|} \a^{\lambda} \sum_{\eta} (-1)^{|\eta|} \a^{-2|\eta|} s_{^{\lambda^t}\!\!/_{\eta}}(q^{-2\rho})  s_{\eta^t} (q^{-2\rho}) \,.
\end{equation}

Finally, when $\a=q^N$ one can show that the above expression without the factor $q^{-\kappa(\lambda)}$
is equal to $\left[\!\!\!\begin{array}{c} N\\ \lambda^t \end{array}\!\!\!\right]$.

%%%%%%%%%%%%%%%%%%%%%%%%%%%%%%%%%%%%%%%%%%%%%%%%%%%%%%%%%%%%%%%%%%%%%%%%%%%

\bibliographystyle{amsalpha}

\end{document}